\documentclass[12pt,english,a4paper]{article}

\pdfoutput=1

\usepackage{amsmath,amsfonts,amssymb,babel,slashed,color,empheq,tensor}
\usepackage{skak}
\usepackage{jheparxiv}
\usepackage[utf8]{inputenc}
\usepackage{amsmath}
\usepackage{amsfonts}
\usepackage{amssymb}
\usepackage{latexsym}
\usepackage{mathrsfs}
\usepackage{braket}		
\usepackage{setspace}
\usepackage{stackengine} 
\usepackage{graphicx}
\usepackage{color}
\usepackage{xcolor}
\usepackage{slashed}
\usepackage{twistor}
\usepackage[all]{xy}
\usepackage{tikz-cd}
\usepackage{float}
\usepackage{roundrule}   
\usepackage{mathtools}
\usepackage{wasysym}
\usepackage{tikz}

\usepackage{scalerel}
\def\rlwd{.9pt}
\def\lhexbrace{\kern1pt%
\setstackgap{S}{0pt}\def\stackalignment{l}
\ThisStyle{\scalerel*{%
  \stackunder[-\rlwd]{%
    \stackon[-\rlwd]{\roundrule{\rlwd}{4pt}}{\rotatebox{60}{\roundrule{4pt}{\rlwd}}}%
  }{\rotatebox{-60}{\roundrule{4pt}{\rlwd}}}%
}{\SavedStyle[}}}
\def\rhexbrace{%
\setstackgap{S}{0pt}\def\stackalignment{r}
\ThisStyle{\scalerel*{%
  \stackunder[-\rlwd]{%
    \stackon[-\rlwd]{\roundrule{\rlwd}{4pt}}{\rotatebox{-60}{\roundrule{4pt}{\rlwd}}}%
  }{\rotatebox{60}{\roundrule{4pt}{\rlwd}}}%
}{\SavedStyle[}}\kern1pt}

\makeatletter

\newcommand{\ib}{\boldsymbol{\mathtt{I}}}
\newcommand{\jb}{\boldsymbol{\mathtt{J}}}
\newcommand{\kb}{\boldsymbol{\mathtt{K}}}
\newcommand{\lb}{\boldsymbol{\mathtt{L}}}
\newcommand{\mb}{\boldsymbol{\mathtt{M}}}
\newcommand{\nb}{\boldsymbol{\mathtt{N}}}
\newcommand{\sG}{\boldsymbol{ \mathsf{G}}}
\newcommand{\cR}{\mathcal{R}}

\newcommand{\mg}{\mathfrak{g}}

\newcommand{\gp}{\tilde{g}_p}

\newcommand{\cQ}{{\cal Q}}

\newcommand{\ttb}{\underline{\mathtt{t}}}

\newcommand{\Ibold}{\boldsymbol{I}}
\newcommand{\Jbold}{\boldsymbol{J}}

\newcommand{\sQ}{\mathsf{Q}}

\newcommand{\mso}{\mathfrak{so}}

\newcommand{\cT}{\mathcal{T}}
\newcommand{\Tr}{\text{Tr}}

\newcommand{\boldJ}{\boldsymbol J}

\newcommand{\diff}{\textnormal{d}}

\usepackage{graphicx}

\newcommand{\nn}{\nonumber}

 \def\one{\mbox{1 \kern-.59em {\rm l}}}

\newcommand{\cK}{\mathcal{K}}

\newcommand{\eps}{\epsilon}

\newcommand{\msf}[1]{\mathsf{\newcommand{\hs}{\mathfrak{hs}}#1}}

\newcommand{\sH}{\mathsf{H}}

\newcommand{\sX}{\mathsf{X}}

\newcommand{\msu}{\mathfrak{su}}

\newcommand{\tJ}{\mathtt{J}}

\def\a{\alpha}  \def\b{\beta}

 \def\L{\Lambda}  
\newcommand{\hsikkt}{$\mathfrak{hs}\text{-IKKT}$}

\makeatother

\begin{document}

 \begin{flushright}
  UWThPh-2024-18 
 \end{flushright}
 \vspace{-10mm}
 
\title{\Large $\mathfrak{hs}$-extended gravity from the IKKT matrix model}

\renewcommand*{\thefootnote}{\fnsymbol{footnote}}
\author[1]{Alessandro Manta,}
\emailAdd{alessandro.manta@univie.ac.at}
\author[1]{Harold C. Steinacker,}
\affiliation[1]{Department of Physics, University of Vienna, \\
Boltzmanngasse 5, A-1090 Vienna, Austria}

\emailAdd{harold.steinacker@univie.ac.at}

\author[1]{Tung Tran\footnote{In transition period to the Asia Pacific Center for Theoretical Physics, Korea.}}

\emailAdd{tung.tran@univie.ac.at}

\abstract{We elaborate further on the one-loop effective action of the IKKT model on $3+1$ dimensional covariant quantum spacetime in the presence of fuzzy extra dimensions. 
In particular,  we describe the one-loop effective action 
in terms of a remarkable $SO(1,9)$ character, which  
allows to evaluate the pertinent traces over the internal modes explicitly. This also allows to estimate the higher-order contributions (in the internal flux $\cF_{\ib\jb}$) to the one-loop effective action in a systematic way. We show that 
all higher-order contributions 
are generally suppressed and UV finite, which justifies the previous 
treatment of the 
induced gravitational action.
We also obtain explicit expressions 
for the effective Newton constant, and determine the dynamics of the Kaluza-Klein scale $\Delta_{\cK}$ of the fuzzy extra dimensions $\cK$. }

\maketitle

\section{Introduction}
\renewcommand*{\thefootnote}{\arabic{footnote}}
\setcounter{footnote}{0}
The generally accepted and well-tested classical description of gravity is provided by General Relativity (GR). However, this theory suffers from classical inconsistencies due to space-time singularities \cite{Penrose:1969pc}, while its incorporation into a consistent quantum theory is obscure \cite{Goroff:1985th}. 

One of the proposals for an underlying quantum gravitational theory is provided by the IKKT matrix model in 0 dimension 
\cite{Ishibashi:1996xs}, which can be viewed as a constructive definition of  type IIB superstring theory in $9+1$ dimensions. A priori, it is not evident how classical $(3+1)$-dimensional space-time along with gravity can emerge from this model. Ideally, this question should be addressed at the non-perturbative level, which is very challenging due to sign problems  \cite{Kim:2011cr,Ito:2015mxa,Nishimura:2019qal}; 
however, important recent progress  \cite{Asano:2024def,Hirasawa:2024dht} should allow to overcome this issue.
We will take a different approach (following \cite{Steinacker:2024unq}) in this work. In particular, we will assume some interesting non-trivial background brane or ``vacuum" of the model which describes our space-time, and study the resulting physics at weak coupling.

One specific mechanism to obtain gravity in the IKKT model was  exhibited in \cite{Steinacker:2021yxt,Steinacker:2023myp}, where it was shown that the Einstein-Hilbert action -- extended by extra terms and degrees of freedom -- can arise from  the one-loop effective action of the IKKT matrix model on some generic $(3+1)$-dimensional space-time branes $M^{1,3}$ in the presence of the fuzzy extra dimensions $\cK$. This mechanism is along the line of the proposal by Sakharov \cite{Sakharov:1967pk}, where the one loop contribution is presumed to be dominant and gravity is viewed as an ``emergence'' effect from a non-gravitational model (see \cite{Visser:2002ew} for a modern review). In the present setup, this idea can thrive due to
various special properties of the model, leading to intricate UV-finite self-interactions of the background branes.
One crucial ingredient is the extra dimensions $\cK$ containing some quantized compact symplectic branes embedded in the transverse directions. Note that these branes should {\em not} be viewed as some result of target space compactification, in contrast to string theory.
At weak coupling, the IKKT matrix model leads to an emergent $(3+1)$-dimensional gauge theory on $M^{1,3}$, which includes an induced Einstein-Hilbert term in the one-loop effective action\footnote{This is also of interest in the context of related work such as  \cite{Brahma:2021tkh}, where the emergence of gravity from the matrix model is simply assumed. Note that the present approach has nothing to do with holography.}.

In a recent series of papers, we considered $M^{1,3}$ to be a covariant quantum space-time $\cM^{1,3}_{\tJ}$ describing a $k=-1$ FLRW cosmological spacetime $\cM^{1,3}$ \cite{Sperling:2018xrm,Sperling:2019xar}.
This background leads to a tower of higher-spin ($\hs$) modes, including the spin-2 field, which describes gravity.
The resulting $\hs$ gauge theory has several peculiar features, 
see e.g. \cite{Fredenhagen:2021bnw,Battista:2023glw,Kumar:2023bxg,Steinacker:2023cuf,Steinacker:2023ntw,Steinacker:2024huv} for recent studies. In particular, it is free of ghosts (negative norm states) \cite{Steinacker:2019awe}, at the expense of manifest Lorentz invariance and being slightly massive. Lorentz invariance, however, is expected to be recovered from the underlying gauge symmetries, which includes $\hs$ volume-preserving diffeomorphisms. Note that the resulting gauge theory is expected to be mildly non-local, in a way which remains to be fully understood.

In the present paper, we focus on the induced gravitational sector of this theory. Building on previous work, we aim to determine the induced gravitational action on the 
background brane $\cM^{1,3}_{\tJ}\times \cK$ in more detail. In particular, we shall apply the geometric trace techniques developed in \cite{Steinacker:2024huv} to do computations related to the internal space $\cK$. We focus on two aspects:
\begin{itemize}
    \item[1-] Refining the approach in \cite{Steinacker:2023myp}, we compute 
    the effective Newton coupling constant  in more detail at one loop. Starting with a general formula based on the $SO(1,9)$ group character, we obtain an explicit geometrical formula
    for the Newton constant, using the geometric trace technique. 

    \item[2-] The one-loop  effective potential governing the fuzzy extra dimensions $\cK$ is computed. In particular, we 
    refine the result of \cite{Steinacker:2024huv} on the stabilization of the radius of $\cK$, and determine the effective potential explicitly at one loop. We show that this  potential has a stable non-trivial minimum  under certain assumptions on the background, which is characterized by two integers $\tJ$ (the higher-spin cutoff) and $d_\cK=\text{dim}\cH_{\cK}$.
  If these integers are very large, the resulting non-trivial vacuum will introduce a mass scale $m_\cK$ to the model and generate 
  a large hierarchy between the IR and UV scales.
    
\end{itemize}

As alluded to in the above, since $\cK$ introduces a scale into the model, the conformal symmetry of the ``would-be" $\cN=4$ SYM sector of the theory is broken by the non-trivial vacuum. The 
proposed background hence provides a mechanism for 
creating large hierarchies through large integers, which 
should be a crucial feature for any model of fundamental physics.

Moreover, 
the aforementioned $SO(1,9)$ character allows us to show that the higher-order corrections in the internal fluxes $\cF_{\ib\jb}$
associated to the above results are generally sub-leading. Similarly, higher-derivative terms in the induced gravitational action are also shown to be suppressed by the UV scale $\Delta_\cK$ following power laws. 
As a result, these contributions are insignificant at low energies.

This paper is organized as follows. Section \ref{sec:2} provides a review of the ($\hs$-)IKKT matrix model on the background brane $\cM^{1,3}_{\tJ}\times \cK$. Section \ref{sec:4} elaborates the study of one-loop effective action in \cite{Steinacker:2023myp}. Section \ref{sec:stabilize} studies the stabilization of $\cK$ and the dynamic of the non-commutativity scale $\Delta_{\cK}$. Section \ref{sec:higher-derivatives} discusses the UV finiteness of the model on the proposed background. We conclude and give some further remarks in Section \ref{sec:discussion}. Useful relations for computation throughout the main text can be found in the appendices.

\section{Review}\label{sec:2}

This section provides short introduction to the higher-spin theory induced by the IKKT matrix model dubbed \hsikkt\ on the $\cM^{1,3}\times \cK$ background brane. We will be rather concise and refer the readers to \cite{Steinacker:2024huv} for further details. 
\paragraph{The IKKT matrix model.} The IKKT matrix model \cite{Ishibashi:1996xs} is defined by a remarkably simple $SO(1,9)$-invariant action 
\begin{align}\label{eq:SO(1,9)action}
    S=\frac{1}{g^2}\Tr\Big([T^{\Ibold},T^{\Jbold}][T_{\Ibold},T_{\Jbold}]+\bar{\Psi}_{\cA}(\Gamma^{\Ibold})^{\cA}{}_{\cB}[T_{\Ibold},\Psi^{\cB}]\Big)\,,\qquad \Ibold=0,1,\ldots,9 \ .
\end{align}
Even though this action is so simple, it is powerful enough to capture the  dynamics of spacetime, as well as emergent fields from matrix degrees of freedom (dof.). Here, $T^{\Ibold}$ are ten $N\times N$ hermitian matrices, and $\Psi$ are $\mso(1,9)$ Majorana-Weyl spinors that act on some Hilbert space $\cH$. Note that $g$ is the coupling of the model, which has dimension $[mass^2]$. Furthermore, the action enjoys the gauge symmetry $\delta T^{\Ibold}=[T^{\Ibold},\xi]$ where $\xi\in \End(\cH)$.

\paragraph{$\cM^{1,3}_{\tJ}$
background brane.} Although there are many solutions in the IKKT matrix model, we will consider backgrounds or ``branes" which can be interpreted in terms of quantized symplectic geometry, as reviewed in e.g. \cite{Steinacker:2019fcb,Steinacker:2024unq}.
More specifically,
we will mostly focus on space-time branes which are sphere bundles over space-time $\cM^{1,3}$, i.e.
\begin{align}
\label{spacetime-brane-S2}
    \cM^{1,3}_{\tJ}\simeq S^2_{\tJ}\times \cM^{1,3} \,  .
\end{align}
Here, $S^2_{\tJ}$ denotes a fuzzy 2-sphere with cutoff $\tJ$, which is responsible for the  higher-spin structures arising on a $k=-1$ FLRW covariant spacetime $\cM^{1,3}$.
Such covariant quantum spacetime allows avoiding manifest breaking of Lorentz invariance by a $B$ field \cite{Sperling:2019xar}. 


Algebraically, such a background $\cM^{1,3}_{\tJ}$ is defined by the following matrix configuration 
\begin{align}\label{eq:background}
T^{\dot\mu} = \frac{1}{R} \cM^{\dot\mu 4}, \qquad \dot\mu = 0,\ldots,3
\end{align}
where $\cM^{AB}$ with $A,B=0,\ldots,5$ are $\mso(2,4)$ generators in the usual unitary 
``doubleton'' minimal representation $\cH_\tJ$ of $SO(2,4)$, cf. \cite{Gunaydin:1998sw}, obeying
\begin{align}
    [\cM^{AB},\cM^{CD}]=\im\Big(\cM^{AD}\eta^{BC}-\cM^{AC}\eta^{BD}-\cM^{BD}\eta^{AC}+\cM^{BC}\eta^{AD}\Big)
\end{align}
as well as further constraints 
discussed in \cite{Sperling:2019xar}. Here, $R$ is the radius of an underlying 4-hyperboloid $H^4$, which will be explained further below. The geometrical significance of these matrices or operators can be understood using  coherent states, which are naturally defined as $SO(1,4)$ group orbit $\ket\zeta = G_\zeta \cdot \ket{0}$ of the lowest weight state $\ket{0} \in \cH_\tJ$. 
These allow to associate functions to operators via the following map\footnote{These coherent states  
satisfy a completeness relation, which will play an important role in the following. }
\begin{align}
\label{symbol-map}
    \End(\cH_\tJ) &\to C^{\infty}(\cM^{1,3}_\tJ),   \nn\\
    \Phi &\mapsto \phi(\zeta) = \langle \zeta|\Phi|\zeta\rangle
\end{align}
 which respects $SO(1,4)$ and provides a faithful description in the semi-classical or IR regime.
In particular, 
the expectation values 
\begin{align}
    x^{\mu}=\langle \zeta|X^{\mu}|\zeta\rangle\,,\qquad X^\mu := \ell_p \cM^{\mu 5}
\end{align}
 where $\ell_p$ is a natural length scale, can be interpreted as coordinate functions on the spacetime brane $\cM^{1,3}\subset \R^{1,3}$, whose 
 length scales 
 obey the following relation:
\begin{align}
\frac{2R}{\ell_p}  \approx \tJ \ \in \N\  .
\end{align}
These space-time coordinates can alternatively be parametrized by hyperbolic coordinates
\begin{align}
    \begin{pmatrix}
     x^0\\
     x^1\\
     x^2\\
     x^3
    \end{pmatrix}=R\cosh(\tau)\begin{pmatrix}
     \cosh(\chi)\\
     \sinh(\chi)\sin(\theta)\cos(\varphi)\\
     \sinh(\chi)\sin(\theta)\sin(\varphi)\\
     \sinh(\chi)\cos(\theta)
    \end{pmatrix}\,.
\end{align}
where $\tau$ will be recognized as a time-like parameter
on the resulting FLRW geometry.

For $\tJ > 0$, the coherent states $\ket\zeta$ are easily seen to sweep out a 6-dimensional space $S^2_\tJ\times H^4$ with an additional ``internal" sphere $S^2_\tJ$, as alluded to in the above. 
In this way,
$\cM^{1,3}_\tJ$ is recognized as a 6-dimensional symplectic space  \eqref{spacetime-brane-S2} which is obtained by a projection of the above sphere bundle, cf. \cite{Sperling:2018xrm,Sperling:2019xar,Steinacker:2023ntw}, or equivalently as the non-compact twistor space $\P^{1,2}$. 
For more details, see e.g. \cite{Steinacker:2024unq}.


\paragraph{$\cM^{1,3}_{\tJ}\times \cK$ background.} 

 To obtain interesting physics including gravity, 
 we will consider backgrounds with product structure 
\begin{align}
    \cM^{1,3}_\tJ\times \cK
\end{align}
 where 
$\cM^{1,3}_\tJ$ describes physical space-time as above, and
$\cK$ describes some compact extra dimensions. Algebraically, such backgrounds are defined by the 
 matrix configurations 
\begin{align}\label{eq:background-2}
T^{\Ibold}=
\binom{T^{\dot\mu}\otimes \one_\cK}{\one_{\cM^{1,3}}\otimes T^{\ib}}
   \,,\qquad  \dot\mu= 0,1, 2, 3\,,\qquad \ib=4,\ldots,9\,
\end{align}
acting on $\cH = \cH_\tJ \otimes \cH_\cK$. Here $T^{\ib}, \ \ib=4,\ldots,9$ describes a compact fuzzy space $\cK$ embedded in the 6 transverse directions. In the semi-classical regime, the $T^{\ib}\sim t^{\ib} =:
\ell_p^{-2} y^{\ib}$ can be viewed as coordinate functions\footnote{Note that all matrices 
$T^{\Ibold}$ are considered to have dimension $[mass]$.} 
on $\cK$. This is in contrast to the space-time brane $\cM^{1,3}_\tJ$, 
where the coordinate functions $x^\mu$ are distinct from the background 
``momentum" generators $T^{\dot\mu} \sim t^{\dot\mu}$ cf. \eqref{eq:background}, which also serve to describe the internal $S^2_{\tJ}$
of $\cM^{1,3}_\tJ$.

Putting this together, the  background brane is described by 9+1
 coordinate functions $x^{\Ibold}:\ \cM^{1,3}\times \cK\xhookrightarrow{}\R^{1,9}$, which are split as
\begin{align}\label{eq:coordinate-split}
    x^{\Ibold}=(x^{\mu},y^{\ib})\,,\qquad \qquad \mu=0,1,2,3\,,\qquad \ib=4,\ldots,9\,.
\end{align}


\paragraph{Quantization maps and space of functions.}

To make sense of the quantum theory, we need to describe the space of functions or harmonics in more detail. 
Since the group orbit of 
$SO(1,4)$ acting on $\ket{0} \in \cH_\tJ$ 
coincides with the non-compact twistor space $\P^{1,2}$ underlying $\cM^{1,3}_\tJ$, we can relate the space of operators $\End(\cH_\tJ)$
with the space of functions
\begin{align}
    C^{\infty}(\cM^{1,3}_\tJ) \cong C^{\infty}(S_{\tJ}^2\times\cM^{1,3})\,.
\end{align}
More explicitly, the coherent states $\ket{\zeta}=G_{\zeta}\cdot |0\rangle$, where $G_{\zeta}\in SO(1,4)$, allow writing down a quantization map 
\begin{align}
\label{quantiz-map}
  \cQ: \ \  C^{\infty}(S^2_{\tJ}\times\cM^{1,3} ) &\to \End(\cH_\tJ),   \nn\\
    \phi(\zeta) &\mapsto \Phi = \int\limits_{S_{\tJ}^2\times\cM^{1,3}} \Omega \,\phi(\zeta) |\zeta\rangle\langle\zeta|
\end{align}
which respects $SO(1,4)$ invariance;
here $\Omega$ is a suitable symplectic measure on $S^2_{\tJ}\times \cM^{1,3}$. This  provides an approximate inverse to \eqref{symbol-map} in the semi-classical regime.
However there is a subtlety concerning
 the spherical harmonics $Y^{sm}$ on the internal $S^2_{\tJ}$. It turns out \cite{Sperling:2018xrm} that the image of $\cQ$  -- i.e. the semi-classical sector of the theory --contains  only a finite tower of modes on $S^2_{\tJ}$, which is truncated at $s=\tJ$.
 This is the reason for identifying $S^2_\tJ$ with a fuzzy sphere\footnote{Strictly speaking, this identification holds only for the modes, but not for the commutation relations. 
The truncation can be verified by acting with a spin Casimir ${\cal S}^2$ on the rhs. of \eqref{quantiz-map}, cf. \cite{Sperling:2018xrm}.}.
 However, $\cQ$  does not cover the full operator algebra 
 $\End(\cH_{\tJ})$, which contains operators with arbitrary spin. 

An analogous statement for the trace will play an important role below.
This truncation of spin can be modeled naturally by the ``local" decomposition\footnote{Such a local product structure can be seen easily from the $H^4_\tJ$ point of view, cf. \cite{Sperling:2018xrm}, or by identifying $\cH_{\tJ}$ as the Hilbert space of a conformal spin $\tJ$ field on $\R^{1,3}$. 
}
\begin{align}
    \cH_\tJ \cong \C^\tJ\otimes\cH_0 
\end{align}
for $\tJ \gg 1$. Note that $\cH_0$ is identified as the Hilbert space associated to $\cM^{1,3}$. Accordingly,  all fluctuations on the present background will take values in a higher-spin algebra $\hs$:
\begin{align}
\End(\cH_\tJ) &\cong \End(\cH_0) \otimes \hs,  \qquad \quad
\hs =\End(\C^{\tJ})\simeq \msu(\tJ)\,,
\end{align}
where $\C^{\tJ}$ is the Hilbert space associated to $S^2_{\tJ}$. 
Here and in the following, we will assume that  
$\tJ \gg 1$ is a large number so that there will be no distinction 
between $\tJ$ and $\tJ+1$.

Since we consider the background with the product structure $\cM^{1,3}_{\tJ}\times \cK$, the Hilbert space $\cH$ and its operator algebra $\End(\cH)$ can be decomposed as
\begin{subequations}
    \begin{align}
    \cH&\cong \cH_\tJ\otimes \cH_{\cK}\,,\\
\End(\cH)&\cong\End(\cH_\tJ)\otimes \mg, 
\end{align}
\end{subequations}
Here, $\cK$ is a quantized symplectic space, whose algebra of functions can be identified with
\begin{align}
    \mg := \End(\cH_\cK)\cong\msu(d_{\cK})
    \,,\qquad d_\cK := \dim \cH_\cK \ < \infty\,.
\end{align}
Note that 
the assumption $d_{\cK}<\infty$ is justified since $\cK$ is compact.
In the semi-classical regime, we can then write 
\begin{subequations}
    \begin{align}
    \cH&\cong \cH_0\otimes\C^{\tJ} \otimes \cH_{\cK}\,,\\
    \End(\cH)&
    \cong C^{\infty}(\cM^{1,3})\otimes \hs \otimes \mg\,.
\end{align}
\end{subequations}
This will be used in the loop computations below.


In general, the above quantization maps provide a faithful description only in the semi-classical, IR regime \cite{Steinacker:2024unq}.
In the one-loop computations, we will only integrate over modes in the image of $\cQ$; this is justified to some extent by the fact that all one loops are UV finite in the maximally supersymmetric model under consideration. 

\paragraph{Local frames.} In the semi-classical limit, the commutator $[\,,]$ decomposes as 
\begin{align}\label{eq:quantum-bracket}
    [a,b]=\im \{a,b\}+[a,b]_{\mg}
\end{align}
where $\{\,,\}$ denotes the Poisson bracket on $\cM^{1,3}_\tJ$ and $[\,,]_{\mg}$ can be viewed as a Lie bracket associated to $\mg$. 
We will typically drop the subscript $\mg$ since it should be clear from the context. 

Now, the Poisson brackets between the (semi-classical) matrix background $t^{\dot\mu}$ and $x^{\nu}$ define a frame
\begin{align}\label{eq:frame}
    E^{\dot\mu\nu} := \{t^{\dot\mu},x^{\nu}\}
\end{align}
for $\dot\mu=\dot 0,\dot 1,\dot 2,\dot 3$ and $\nu=0,1,2,3$. 
Here, $t^{\dot\mu}$ is assumed to be a mild deformation of the undeformed background $\ttb^{\dot\mu}$, cf.  
\eqref{eq:background}, with
\begin{align}\label{eq:PoissonM13}
    E^{\dot\mu\nu} := \{\ttb^{\dot\mu},x^{\nu}\}&=\frac{x^4}{R}\eta^{\dot\mu\nu}=\sinh(\tau)\eta^{\dot\mu\nu}\,,\qquad \eta^{\dot\mu\nu}=\diag(-,+,+,+)\,,
\end{align}
This frame can be used to obtain the \emph{auxiliary} metric $\gamma_{\mu\nu}$ and the \emph{effective} metric $G_{\mu\nu}$ as 
\begin{align}\label{eq:eff-G}
    G_{\mu\nu} 
    =\rho^2\gamma_{\mu\nu} \,,\qquad \gamma_{\mu\nu}:=E^{\dot\kappa}{}_{\mu}E_{\dot\kappa\nu}\,,
\end{align}
which for the undeformed background are given by 
\begin{align}
    G_{\mu\nu} = \sinh(\tau) \eta_{\mu\nu}
    \,,\qquad \rho^2=\sinh^3(\tau)\,,\qquad \gamma_{\mu\nu}:=\sinh^{-2}(\tau)\eta_{\mu\nu}\,.
\end{align}
Note that the \emph{auxiliary} metric $\gamma_{\mu\nu}$ is similar to the open string metric in string theory, while $\rho$ can be interpreted as a dilaton. 
Using the frame \eqref{eq:frame}, we can always trade a pair of contracted frame (dotted) indices for a pair of contracted spacetime (undotted) indices, for instance, as
\begin{align}\label{eq:converting-rule}
    A^{\dot\mu}B_{\dot\mu}=E^{\dot\mu\mu}A_{\mu}E_{\dot\mu}{}^{\nu}B_{\nu}=\gamma^{\mu\nu}A_{\mu}B_{\nu}=\rho^2G^{\mu\nu}A_{\mu}B_{\nu}\,.
\end{align}
Due to the fact that we have two different metrics, i.e. auxiliary and effective cf. \eqref{eq:eff-G}, we should be very careful when rising and lowering the spacetime indices. Thus, to avoid confusion, we will spell out the appropriate metrics whenever it is necessary.




\paragraph{Pertinent traces.}

We will need two different types of traces in the following:


$\bullet$ \underline{Trace over $\cH$ as integral over $\cM^{1,3}_{\tJ}\times \cK$.} This is the trace used in the semi-classical regime where we can 
relate the trace over $\cH$ to an integral over the semi-classical functions as follows.
Denoting $\tr_{\cH}:=\tr_{\cH_{0}}\times \tr_{\C^{\tJ}}\times\tr_{\cH_{\cK}}$ for the local trace over $\cH$, we have \cite{Steinacker:2024huv}
\begin{align}\label{eq:classical-trace}
    \tr\,\cO=\int\limits_{\cM^{1,3}} \mho_0\int_{S^2_{\tJ}} \frac{\varpi}{2\pi}\,\tr_{\cH_{\cK}}\cO\,,\qquad \mho_0=\frac{R}{\ell_p^4\, x_4}d^4x\,,\quad \cO\in \End(\cH_\tJ)\otimes \mg\,.
\end{align}
Here we have identified the operator $\cO$ on the lhs. 
with a $\mg$-valued semi-classical  function on $\cM^{1,3}_{\tJ}$ on the rhs. using the symbol map \eqref{symbol-map}.
Here, $\mho_0$ is the symplectic volume form on the undeformed $\cM^{1,3}$, and $\ell_p$ is a natural length scale introduced to balance the dimension of the measure $d^4x$.
Furthermore,  $\varpi$ is the volume form of $S^2_{\tJ}$ normalized as
\begin{align}\label{eq:varpi-normalize}
    \int_{S^2_{\tJ}}\frac{\varpi}{2\pi}=\tJ\,.
\end{align}

$\bullet$ \underline{Trace over 
$\End(\cH)$ as integral over semi-classical functions.} For the loop computations in the quantum regime, we need to evaluate the trace 
\begin{align}\label{eq:quantum-trace}
    \Tr_{\End(\cH)} = \Tr_{ \End(\cH_{\tJ})\otimes \End(\cH_\cK)}=\Tr_{\End(\cH_{0})}\times \Tr_{\hs}\times \Tr_{\mg}
\end{align}
over the space of operators $\End(\cH)$. In general, this would be very non-local and hard to evaluate. In the present context, the traces will be UV finite, so that it makes sense to reduce the space of operators to the semi-classical regime, i.e. the image of the quantization map \eqref{quantiz-map}.
Then, the space $\End(\cH_\tJ)$ can be described as
\begin{align}
\label{hs-modes-explicit}
\Phi = \sum_{s=0}^\tJ \sum_{m=- s}^s\phi_{sm}\hat Y^{sm} = \cQ(\phi)\,,
\end{align}
where 
\begin{align}
    \phi = \sum_{s=0}^\tJ\sum_{m=-s}^s \phi_{sm}(x) Y^{sm}\,,\qquad \phi_{sm}(x)\in C^{\infty}(\cM^{1,3})\,.
\end{align}
Hence, the trace over 
$\End(\cH_0)$ is equivalent to a (convergent)  integral over
$C^{\infty}(\cM^{1,3})$, which is evaluated in the over-complete basis of normalized Gaussian wave packets $\psi_{k;x}$ localized around a point $x\in \cM^{1,3}$ with the characteristic wave vector $k^{\mu}$ (see \cite{Steinacker:2023myp,Steinacker:2024unq} for more detail).
 Then, for a generic  operator $\cO$ acting on $\End(\cH_{\tJ})\otimes \mg$, we have \cite{Steinacker:2023myp}
\begin{align}
    \Tr\,\cO=\int\limits_{\cM^{1,3}} d^4x\sqrt{G}\int \frac{d^4k}{(2\pi)^4\sqrt{G}}\Tr_{\hs}\Big(\Tr_{ \mg}\langle\psi_{k,x},\cO\psi_{k,x}\rangle\Big)\,
\end{align}
where $\Tr_\hs$ indicates the sum over all higher-spin modes encoded by the pair $(s,m)$. Note that we will systematically drop the absolute value notation in $\sqrt{|G|}$ since no confusion can arise.

In the presence of a non-trivial $\cK$ space in extra dimensions, 
the trace over $\mg$ can be evaluated using string modes on $\cK$, which are (possibly non-local) generalizations of Gaussian wave packets. For simplicity, we will assume $\cK$ is irreducible but not a point brane, so that interesting physics can emerge. Then,
\begin{align}\label{eq:geo-trace-K}
    \Tr_{\mg}\cO(x,y)=\int\limits_{\cK\times \cK}\frac{\Omega_x\times\Omega_y}{(2\pi)^{|\cK|}}\left(^x_{y}\right|\cO \left|^x_{y} \right)\,,\qquad \cO(x,y)\in \cH_{\cK,x}\otimes \cH_{\cK,y}^*
\end{align}
where $(x,y)$ are the end points of the open string modes $\left|^x_{y} \right)=|x)(y|$ on $\cK$.


\paragraph{Kaluza-Klein and higher-spin mass scales.} As illustrated in \cite{Steinacker:2021yxt,Steinacker:2023myp,Battista:2023glw,Kumar:2023bxg,Steinacker:2024huv}, certain interesting low-energy physics can arise from the IKKT matrix model in the presence of extra dimensions $\cK$. This compact fuzzy space introduces two crucial scales to the model:  the non-commutativity scale $\Delta_{\cK}$ 
on $\cK$ defined by $\Delta^2_{\cK} = \cO(\cF^{\ib\jb})$, and
the KK scale $m_{\cK}$ which characterizes the maximal eigenvalue of $\Box_6:=[\ttb^{\ib},[\ttb_{\ib},.]]$, i.e. the UV cutoff scale on $\cK$. Here, $\ttb^{\ib}$ is identified with the embedding  coordinates $z^{\ib}$ via\footnote{Note that $z^{\ib}$ are dimensionless rescalings of $y^{\ib}$, cf.  \eqref{eq:coordinate-split}.}
\begin{align}\label{t-z}
    \ttb^{\ib} = m_{\cK}\,z^{\ib} : \ \cK \hookrightarrow \R^6\,,\qquad z^{\ib} z_{\ib} = \cO(1)\,,\qquad \ib=4,\ldots,9\,.
\end{align}
Note that $\Delta_{\cK}$ is related to $m_{\cK}$ via the Bohr-Sommerfeld quantization condition \cite{Steinacker:2024huv} 
\begin{align}\label{eq:Bohr-Q}
    d_{\cK}=\int \frac{\Omega_{\cK}}{(2\pi)^{|\cK|/2}}
\end{align}
as
\begin{align}\label{dK-mK}
d_\cK \Delta_\cK^{|\cK|} \approx m_\cK^{|\cK|} \qquad \text{or equivalently} \qquad \Delta_{\cK}\approx\frac{1}{d_{\cK}^{1/|\cK|}}m_{\cK}\,.
\end{align}
Besides $\Delta_{\cK}$, we will also encounter the $\hs$ mass scale $\Delta_{\tJ}$ associated with $S^2_{\tJ}$. This arises from acting with the $(3+1)$-dimensional matrix Laplacian $\Box_{1,3}:=[t^{\dot\mu},[t_{\dot\mu},-]]$ on a $\hs$-valued operator $\Phi = \phi_{sm}(x) \hat Y^{sm}$ (cf. \eqref{hs-modes-explicit}): 
\begin{align}\label{eq:3+1dBox}
\Box_{1,3} (\phi_{sm}(x) \hat Y^{sm}) =\hat Y^{sm}(\Box_{1,3}+m_s^2)\phi_{sm}(x) \,.
\end{align}
For the undeformed $\cM^{1,3}_\tJ$ background, one finds 
$m^2_s = 
-\frac{s}{R^2}$ (
taking into account appropriate factors of $\rho^2$
), which 
amounts to an IR mass scale set by the cosmic curvature, cf. \cite{Steinacker:2023cuf}.
However, that specific value should be taken with a grain of salt, since it depends on the details of the background. 
In particular, the negative sign  is presumably an artifact, which is expected to be cured by considering a more general background as in \cite{Battista:2023glw}. 
Note that as long as $m_s$ 
is in the IR (cosmological) regime, 
there will be no serious consistency problem.\footnote{This negative mass reflects the fact that the background 
$T^{\dot\mu} = t^\mu$ 
defining $\cM^{1,3}$ is not fully consistent recalling that $\Box_{1,3} T^{\dot\mu}  = \frac{3}{R^2} T^{\dot\mu} \neq 0$, cf. \cite{Steinacker:2024unq}. } In particular, we show in Appendix \ref{app:verify} that if the background $T^{\dot\mu}$ is modified to
\begin{align}
    T^{\dot\mu}=f(\tau)\ttb^{\dot\mu}\,,
\end{align}
then for some suitable function $f(\tau)$, which preserves $SO(1,3)$ (cf. \cite{Battista:2023glw}), the mass $m_s^2$ can be positive and time-dependent.
We will, therefore, replace $m_s$ by some generic positive $\hs$ scale, i.e.
\begin{align}
    m_s^2 \ \to \ \Delta_\tJ^2 \ > 0 \ .
\end{align}
This replacement will allow us to simplify the sum over $\hs$ modes in the 1-loop computations. 
Furthermore, to ensure a large hierarchy between KK and $\hs$ scales\footnote{Note that this hierarchy requirement may not be essential; the only essential requirement is that the gravity modes are (almost) massless.}, we will typically assume 
\begin{align}
    \Delta_{\cK}\gg \Delta_{\tJ}  \ .
    \label{DeltaK-J-hierarchy}
\end{align}
This will also play a role in the discussion of the stability of $\cK$ in Section \ref{sec:stabilize}. 




\paragraph{Physical mass scales.} Let us now discuss a more subtle issue regarding the evolution of the above mass scales. Due to the $4+6$ splitting of the product structure $\cM^{1,3}_{\tJ}\times \cK$ and taking into account \eqref{eq:3+1dBox},  the 10-dimensional matrix Laplacian splits as 
\begin{align}
\label{eq:Box-decomposition}
    \Box:=[\ttb^{\Ibold},[\ttb_{\Ibold},-]] = \Box_{1,3} + \Box_6 \ .
\end{align}
In the semi-classical regime on generic backgrounds $\cM^{1,3}$, 
the space-time operator $\Box_{1,3}$ 
acting on functions $\phi(x)$
takes the following form \cite{Steinacker:2023myp}
\begin{align}\label{eq:BoxG}
    \Box_{1,3}&=-\{t^{\dot\mu},\{t_{\dot\mu},-\}\}=\rho^2\Box_G\,,\qquad \Box_G=
   -\frac{1}{\sqrt{G}}\p_\mu\big (\sqrt{G}\, G^{\mu\nu}\p_\nu) \ .
\end{align}
Here, $\Box_G$ is the standard d'Alembertian associated with the effective metric $G_{\mu\nu}$. On the other hand, when acting on a KK eigenmode $\Upsilon_{\Lambda}$, $\Box_6$ acquires the following eigenvalue
\begin{align}
    \Box_6 \Upsilon_\L = (\mu_\L^2 m_\cK^2) \Upsilon_\L\,,\qquad \mu_\L^2\in \Q^+\quad \text{and} \quad 0\leq\mu_\L\leq 1\,.
\end{align}
Here, we use  $\mu_\L$ is used to describes the discrete spectrum of $\cK$ given that $m_{\cK}$ is the maximal eigenvalue of $\Box_6:=[\ttb^{\ib},[\ttb_{\ib},-]]$. Then, if $\varphi_\L = \phi_\L(x) \Upsilon_\L$ is an eigenmode on $\cM^{1,3}\times \cK$, it will
satisfy the following on-shell relation
\begin{align}\label{eq:effective-evolution}
   0 = \Big(\Box_G+ \rho^{-2}(\mu_\L^2 m_\cK^2 + m_s^2)\Big)\varphi_\L\,
\end{align}
taking into account \eqref{eq:3+1dBox}.
This motivates the following definition of effective KK mass
\begin{align}
\label{evolution-KK-masses}
    m^2_{\L,{\rm eff}} = \mu_\L^2 \rho^{-2} m_\cK^2\,
\end{align}
and similarly the effective $\hs$ mass $m^2_{s,{\rm eff}} = \rho^{-2} m^2_s$.
In particular, these masses evolve with the expansion of the universe with $\rho^{-2}$. 
We will see that the Planck scale 
involves the same $\rho^{-2}m_\cK^2$ factor. This in turn means that the time dependence through $\rho$ drops out if we express the local physics in terms of dimensionless {\em ratios} of these dynamical scales.
Note also that the negative masses $m_s^2$ will either be dropped or replaced by the positive scale $\Delta^2_\tJ$ as discussed above.



\section{Emergent \texorpdfstring{$\hs$}{hs}-extended gravity at one loop}\label{sec:4}
We now elaborate further the  one-loop effective action for the geometric or gravitational sector of \hsikkt \ on the present $\cM^{1,3}_{\tJ}\times \cK$ background, as initiated in \cite{Steinacker:2021yxt,Steinacker:2023myp,Battista:2023glw,Kumar:2023bxg}. In particular, we introduce an $SO(1,9)$ character denoted as $\sQ_{10}$, which provides simple closed expressions for different components of the one-loop effective action. Focusing on the emergent ($\hs$-extended) gravity sector, we show that 
the lowest-order contribution 
in the internal flux $\cF_{\ib\jb}$ 
to the induced gravitational action is indeed dominant, thus justifying the previous treatment 
in
\cite{Steinacker:2021yxt,Steinacker:2023myp}.
The present systematic approach  provides a nice geometric formula for the effective Newton coupling constant.

\subsection{One loop effective action}\label{sec:3.1}
\paragraph{General setup.} To begin the study of the one loop effect action, let us introduce the following $10d$ matrix field strength
\begin{align}
    \cF^{\Ibold\Jbold}:=-\im[t^{\Ibold},t^{\Jbold}]=\begin{pmatrix}
        -\im[t^{\dot\mu},t^{\dot\nu}] & -\im [t^{\dot\mu},t^{\ib}]\\
        -\im [t^{\ib},t^{\dot\mu}]&-\im [t^{\ib},t^{\jb}]
    \end{pmatrix}=\begin{pmatrix}
        \cF^{\dot\mu\dot\nu} & \cF^{\dot\mu\ib}\\
        \cF^{\ib\dot\nu} & \cF^{\ib\jb}
    \end{pmatrix}\,.
\end{align}
Integrating out (after gauge fixing) the fluctuating fields in the Gaussian approximation around some given background 
$\bar T$
with a regulator $\im \epsilon$ term \cite{Ishibashi:1996xs,Blaschke:2011qu,Steinacker:2023myp}, the partition function
\begin{align}\label{eq:measure}
    Z_{\text{1-loop}}=\int dT d\Psi d\overline{\Psi} dcd\bar{c}\,e^{\im S_{\text{reg}}[T,\Psi,\overline{\Psi},c,\bar{c}]}
\end{align}
leads to the following one-loop effective action 
\begin{align}
    S_{\rm eff}[T] = S_0[T]+ \Gamma_{\rm 1-loop}[T]
\end{align}
where $S_0$ denotes the classical piece and 
\small
\begin{align}\label{Gamma-schwinger-susy}
\Gamma_{\rm 1-loop}
&= +\frac \im 2 \Tr_{\End(\cH)\otimes \cR} \Bigg[\!\log\Big(\Box -\im\varepsilon -\Sigma^{(V)}_{\Ibold\boldJ}[\cF^{\Ibold\boldJ},-]\Big)
  - \frac 12 \log\Big(\Box -\im\varepsilon -\Sigma^{(\Psi)}_{\Ibold\boldJ}[\cF^{\Ibold\boldJ},-]\Big) 
 - 2 \log\Big(\Box-\im\varepsilon\Big) \! \Bigg]  \nn\\
 &= -\frac{\im}{2}\int_0^{\infty}\frac{d\alpha}{\alpha}\Tr_{\End(\cH)\otimes \cR}\Bigg[e^{-\im\alpha \,\Box}\Big(e^{\im\alpha\, \Sigma_{\Ibold\Jbold}^{(V)}[\cF^{\Ibold\Jbold},-]}-\frac{1}{2}e^{\im \a\,\Sigma_{\Ibold\Jbold}^{(\Psi)}[\cF^{\Ibold\Jbold},-]}-2\Big)\Bigg]\,.
\end{align}
\normalsize
Here, $\cR$ denotes the representation space that we will trace over, e.g. $(V)$ for vector and $(\Psi)$ for spinor representations.
Note that to go from the first to the second line above, we have used the identity
\begin{align}
\label{duhamel-identity}
\log\frac{Y-\im\varepsilon}{X-\im\varepsilon} \, &= 
\int_0^\infty\frac{d\a}{\a}\Big[e^{-\im\a (X-\im\varepsilon)} - e^{-\im\a (Y-\im\varepsilon)}\Big]\,.
\end{align}
Furthermore the $-\im\varepsilon$ in the exponent on the rhs. of \eqref{duhamel-identity} is consistent with a deformation $\a\to \a -\im \varepsilon$ for the Schwinger parameter $\a$, which will be understood in the following\footnote{This is manifest for positive $X,Y$; negative $X,Y$ will only arise from the time-like momenta, which are consistently Euclideanized by the standard contour prescription.}. 
Here, we have assumed that the background is slowly varying so that
\begin{align}
    [\Box,\cF^{\Ibold\Jbold}]=0\,,\qquad \cF^{\dot\mu\ib}\approx 0\,.
\end{align}


\paragraph{The $SO(1,9)$ character.}

For convenience, let us express $\Gamma_{\rm 1-loop}$ in terms of the following $SO(1,9)$ character
\begin{align}\label{eq:Q10}
    \sQ_{10}:&=\Tr_{\cR}\Big(e^{\im\alpha \,\Sigma_{\Ibold\Jbold}^{(V)}[\cF^{\Ibold\Jbold},-]}-\frac{1}{2}e^{\im\alpha\, \Sigma_{\Ibold\Jbold}^{(\Psi)}[\cF^{\Ibold\Jbold},-]}-2\Big)\nn\\
   & \equiv \Tr_{\cR}\Big(e^{\im\alpha \,\Sigma_{\Ibold\Jbold}^{(V)}\delta\cF^{\Ibold\Jbold}}-\frac{1}{2}e^{\im \alpha\,\Sigma_{\Ibold\Jbold}^{(\Psi)}\delta\cF^{\Ibold\Jbold}}-2\Big)\,,
\end{align}
where $\cR$ denotes the appropriate representation. 
Furthermore, $\delta \cF_{\Ibold\Jbold}$ stands for $[\cF_{\Ibold\Jbold},-]$,
which will become more transparent upon using an open string basis\footnote{In practice, we have e.g. $[\cF,\left|^{x}_{y} \right)]=\cF(x)-\cF(y)\equiv \delta \cF$. See further discussion in \cite{Steinacker:2022kji}.}.
By virtue of the $4+6$ splitting
 of the proposed background $\cM^{1,3}_{\tJ}\times \cK$, we can decompose 
\begin{subequations}
    \begin{align}
        (V) &=(4) \oplus (6)\,,\\
        (\Psi)&= \Big((2_-) \otimes (4_-)\Big)\oplus \Big((2_+) \otimes (4_+)\Big)\,.
    \end{align}
\end{subequations} 
Here, $(4)$ and $(6)$ denote the vector  representations while $(2_\mp)$ and $(4_\mp)$ denote the chiral spinor representations of $SO(1,3)$ and $SO(6)$, respectively. Then, $\sQ_{10}$ can be cast into the following form
\small
\begin{align}\label{eq:Q46}
    \sQ_{10}=&\Tr_{V_{(4)}}\Big(e^{\im\alpha \Sigma_{\dot\mu\dot\nu}^{(4)}\delta\cF^{\dot\mu\dot\nu}}\Big)+\Tr_{V_{(6)}}\Big(e^{\im\alpha \Sigma_{\ib\jb}^{(6)}\delta\cF^{\ib\jb}}\Big)\nn\\
    &\qquad \qquad -\frac{1}{2}\sum_{\mp}\Tr_{\Psi_{(2_\mp)}}\Big(e^{\im \a\,\Sigma_{\dot\mu\dot\nu}^{(2_\mp)}\delta\cF^{\dot\mu\dot\nu}}\Big)\Tr_{\Psi_{(4_\mp)}}\Big(e^{\im\a\, \Sigma_{\ib\jb}^{(4_\mp)}\delta\cF^{\ib\jb}}\Big)-2\,.
\end{align}
\normalsize
From this character, one can extract all possible contributions in the $\a$ expansion as shown below.

\subsection{\texorpdfstring{$\hs$}{hs}-extended gravitational action at one loop}


Let us illustrate the power of $\sQ_{10}$ by focusing on the lowest-derivative contribution to the gravitational action on $\cM^{1,3}$. We will expand the above $SO(1,9)$ character to the first non-trivial order $\delta\cF^2_{\dot\mu\dot\nu}$. 
For convenience, we introduce the following $SO(1,3)$ characters (cf. \cite{Steinacker:2024huv})
\begin{subequations}
    \begin{align}
 \chi_{\cM}^{(4)}[\a]
    &:= \Tr_{V_{(4)}}\Big(e^{\im\alpha \,\Sigma_{\dot\mu\dot\nu}^{(4)}\delta\cF^{\dot\mu\dot\nu}}\Big) 
     =4 -2\a^2\delta\cF_{\dot\mu\dot\nu}\delta\cF^{\dot\mu\dot\nu} + \cO(\a^4)\,,\\
    \chi_{\cM}^{(2_\mp)}[\a] &:=
    \Tr_{\Psi_{(2_\mp)}}\Big(e^{\im \a\,\Sigma_{\dot\mu\dot\nu}^{(2_\mp)}\delta\cF^{\dot\mu\dot\nu}}\Big)
    =2-\frac{\a^2}{2}\delta\cF_{\dot\mu\dot\nu}\delta\cF^{\dot\mu\dot\nu} + \cO(\a^4)\,.
\end{align}
\end{subequations}
These allow us to write the $SO(1,9)$ character as 
\begin{align}
    \sQ_{10} &= \sX_{6}
    + \a^2 \delta\cF_{\dot\mu\dot\nu}\delta\cF^{\dot\mu\dot\nu}\Big(-2 + \frac 14  \sum_\pm \Tr_{\Psi_{(4_\pm)}}\big(e^{\im\a\,\Sigma^{(4_\pm)}_{\ib\jb}\delta\cF^{\ib\jb}}\big)  \Big)
     + \cO(\delta\cF_{\dot\mu\dot\nu}^4) \ .
\end{align}
Here,
\begin{align}
\label{eq:character-K-scalar-UV-local}
 \sX_6[\a]
 &= 2+
\Tr_{V_{(6)}}\big(e^{\im\a\,\Sigma^{(6)}_{\ib\jb}\delta\cF^{\ib\jb}}\big)
 - \sum_\pm \Tr_{\Psi_{(4_\pm)}}\big(e^{\im\a\,\Sigma^{(4_\pm)}_{\ib\jb}\delta\cF^{\ib\jb}}\big) \, \nn\\
 &= \alpha^4\Big(4\delta \cF^{\ib\jb}\delta \cF_{\jb\kb}\delta \cF^{\kb\mb}\delta \cF_{\mb\ib}-\delta \cF^{\ib\jb}\delta \cF_{\ib\jb}\delta \cF^{\mb\nb}\delta\cF_{\mb\nb}\Big)+\cO(\a^6)\,
\end{align}
is a remarkable character of $SO(6)$ introduced in \cite{Steinacker:2024huv}, which governs the effective potential for $\cK$ as discussed below.
Similarly,
\begin{align}
\sG_6[\a]
 :&= -2 + \frac 14  \sum_\pm \Tr_{\Psi_{(4_\pm)}}\big(e^{\im\a\,\Sigma^{(4_\pm)}_{\ib\jb}\delta\cF^{\ib\jb}}\big) 
 = -\frac 32 -\frac 14\sX_6 + \frac 14 \Tr_{V_{(6)}}\big(e^{\im\a\,\Sigma^{(6)}_{\ib\jb}\delta\cF^{\ib\jb}}\big)\nn\\
&= -\frac 12 \a^2 \delta\cF_{\ib\jb}\delta\cF^{\ib\jb} + \cO(\a^4)
\end{align}
is another $SO(6)$ character which governs the induced gravitational (and Yang-Mills \cite{Steinacker:2024huv}) action. To obtain the above results, the following relations are useful
\begin{subequations}\label{eq:various-trace-reps}
    \begin{align}
\Tr_{V_{(4)}}\Big(\Sigma^{V_{(4)}}_{\dot\mu\dot\nu}\Sigma^{V_{(4)}}_{\dot\rho\dot\sigma}\Big) &= 
2\big(\eta_{\dot\mu\dot\rho}\eta_{\dot\nu\dot\sigma} - \eta_{\dot\mu\dot\sigma}\eta_{\dot\nu\dot\rho}\big)\,,\\
\Tr_{V_{(6)}}\Big(\Sigma^{V_{(6)}}_{\ib\jb}\Sigma^{V_{(6)}}_{\mb\nb}\Big)&=2\big(\delta_{\ib\mb}\delta_{\jb\nb}-\delta_{\ib\nb}\delta_{\jb\mb}\big)\,,\\
\Tr_{\Psi_{(2_\mp)}}\Big(\Sigma^{\Psi_{(2_\mp)}}_{\dot\mu\dot\nu}\Sigma^{\Psi_{(2_\mp)}}_{\dot\rho\dot\sigma} \Big)&= \frac 12(\eta_{\dot\mu\dot\rho}\eta_{\dot\nu\dot\sigma} - \eta_{\dot\mu\dot\sigma}\eta_{\dot\nu\dot\rho})\,,\\
\Tr_{\Psi_{(4_\mp)}}\Big(\Sigma^{\Psi_{(4_\mp)}}_{\ib\jb}
\Sigma^{\Psi_{(4_\mp)}}_{\kb\lb}\Big) &= \delta_{\ib\kb}\delta_{\jb\lb} - \delta_{\ib\lb}\delta_{\jb\kb}\,.
    \end{align}
\end{subequations}
Observe that 
the expansion of $\sQ_{10}$ starts at $\cO(\a^4)$ as a result of maximal supersymmetry. This is crucial for the (approximate) locality of the 1-loop effective action.



The higher-spin extended variant of gravity is associated with the following contribution
\begin{align}
    \sQ_{10}\Big|_{\delta\cF^{\dot\mu\dot\nu}\delta \cF_{\dot\mu\dot\nu}}=\sQ_{\textnormal{grav}}\,.
\end{align}
In other words,
\begin{equation}
    \sQ_{\textnormal{grav}}:=\alpha^2\delta \cF_{\dot\mu\dot\nu}\delta\cF^{\dot\mu\dot\nu}\sG_6[\alpha]
\end{equation}
is the character which gives the following one-loop effective action for the $\hs$-extended gravity
\begin{align}
    \Gamma_{\rm 1-loop}^{\textnormal{grav}}=-\frac{\im}{2}\int \frac{d\alpha}{\alpha}\Tr_{\End(\cH_{\cM})\otimes \hs\otimes \mg}\Big[e^{-\im\alpha\,\Box}\sQ_{\textnormal{grav}}\Big]\,.
\end{align}
The key step to obtain the induced gravity action is to evaluate
\begin{align}
    \delta\cF_{\dot\mu\dot\nu}=[\cF_{\dot\mu\dot\nu},-] \sim +\im\,\{\cF_{\dot\mu\dot\nu},-\}\,
\end{align}
as a derivative operator on $\cM^{1,3}$.
This is justified since we can treat $\cF_{\dot\mu\dot\nu}$ as a $\hs$-valued slowly varying field strength, as discussed previously. 
Following the treatment in \cite{Steinacker:2023myp}, we can then write
\begin{align}
    \delta\cF_{\dot\mu\dot\nu}\delta \cF^{\dot\mu\dot\nu}
    &=-\{\cF^{\dot\mu\dot\nu},x^{\alpha}\}\p_{\alpha}(\{\cF_{\dot\mu\dot\nu},x^{\beta}\}\p_{\beta})
    \approx -\cT^{\dot\mu\dot\nu\alpha}\cT_{\dot\mu\dot\nu}{}^{\beta}\p_{\alpha}\p_{\beta}
\end{align}
in terms of the Weitzenböck torsion
\begin{align}
    \cT^{\dot\mu\dot\nu\alpha}:=\{\cF^{\dot\mu\dot\nu},x^{\alpha}\} \ ,
\end{align}
which encodes the non-trivial curvature on the deformed background  $\cM^{1,3}$ \cite{Steinacker:2020xph}.
This is the key ingredient for acquiring the Einstein-Hilbert action\footnote{
In particular, we ignore possible contributions from the $\hs$ generators $\{\cF^{\dot\mu\dot\nu},u^\sigma\}$ under the trace $\Tr_\hs$, because these do not contribute to the torsion and the induced gravitational action. The significance of these terms should be investigated elsewhere, but they appear to be subleading.}. 
Upon evaluating $\delta \cF_{\dot\mu\dot\nu}\delta \cF^{\dot\mu\dot\nu}$ on the (approximate) plane wave basis $\psi_{k,x}$ as in \cite{Steinacker:2023myp}, we obtain
\begin{align}
    \delta\cF_{\dot\mu\dot\nu}\delta \cF^{\dot\mu\dot\nu}\approx+\cT^{\dot\mu\dot\nu\alpha}\cT_{\dot\mu\dot\nu}{}^{\beta}k_{\alpha}k_{\beta}\,.
\end{align}%
Next, recall that $\Box=\Box_{1,3}+\Big(\Box_6+\Delta_{\tJ}^2\Big)$, cf. \eqref{eq:Box-decomposition} where $\Box_{1,3}=\rho^2\Box_G$ cf. \eqref{eq:BoxG} and $\Delta_{\tJ}$ is used to simplify $m^2_s$ \eqref{eq:3+1dBox}. Upon evaluating $\Box_{1,3}$ on plane wave basis in local coordinate where $\p\gamma=0$, we get
\begin{align}
  \Box_{1,3}\mapsto  \tilde k^2\,,\qquad \tilde k^2=k_{\mu}k_{\nu}\gamma^{\mu\nu}=k_{\mu}k_{\nu}\rho^2G^{\mu\nu}\,
\end{align}
At this stage, the integral over $k$'s can be done using the master formula \eqref{eq:general-k-integral}. In particular, for the case in consideration, we have
\begin{align}
    \int \frac{d^4k}{(2\pi)^4\sqrt{G}}k_{\alpha}k_{\beta}e^{-\im \alpha k_{\mu}k_{\nu}\rho^2G^{\mu\nu}}=-\frac{1}{2(4\pi)^2}\frac{1}{\rho^6\alpha^3}G_{\alpha\beta}=-\frac{1}{2(4\pi)^2}\frac{1}{\rho^4\alpha^3}\gamma_{\alpha\beta}\,.
\end{align}
Then, the full one loop effective action for the $\hs$-extended gravity reads
\begin{align}
    \Gamma_{\rm 1-loop}^{\textnormal{grav}}&=+\frac{\im}{4(4\pi)^2}\int \frac{\sqrt{G}}{\rho^4}\Tr_{\hs\otimes \mathfrak{g}}\left(\int\frac{\diff\alpha}{\alpha^2}\e^{-\im\alpha\,(\square_6+\Delta_{\tJ}^2)}\cT^{\dot\mu\dot\nu\alpha}\cT_{\dot\mu\dot\nu}^{\ \ \beta}\gamma_{\alpha\beta}\sG_6[\alpha]\right)\,.
\end{align}
The term in the parentheses contains the inverse Newton gravitational constant, as it is the coupling to the Ricci scalar in the effective action (see e.g. \cite{Steinacker:2023myp}). 
Using the conventions\footnote{To avoid any ambiguities,  the torsion tensor is always written with two lower and one upper index.} that the indices of the torsion tensor are
covariantized in terms of the frame $E^{\dot\a}_\mu$ corresponding to the
metric $\gamma^{\mu\nu} = \rho^2 G^{\mu\nu}$, we can rewrite \cite{Steinacker:2023myp}
\begin{align}
 \tensor{\cT}{^{\dot\mu}^{\dot\nu}^{\a}}\tensor{\cT}{_{\dot\mu}_{\dot\nu}^{\b}} G_{\a\b}
    = \rho^4 \cT^{\mu}{}_{\nu\a}\cT_{\mu\ \ \b}^{\ \nu}G^{\a\b}\ .
    \label{torsion-contraction-frame-G}
\end{align}
Collecting these results, the one loop effective action can be cast into the form
\begin{align}
    \Gamma_{\rm 1-loop}^{\textnormal{grav}}&=+\frac{\im}{4(4\pi)^2}\int \frac{\sqrt{G}}{\rho^2}\Tr_{\hs\otimes \mathfrak{g}}\left(\int\frac{\diff\alpha}{\alpha^2}\e^{-\im\alpha\,(\square_6+\Delta_{\tJ}^2)}\sH_{\textnormal{grav}}\sG_6[\alpha]\right)\,.
\end{align}
where for organizational purpose, we have introduced
\begin{align}
    \sH_{\textnormal{grav}}:= \cT^{\mu}{}_{\nu\a}\cT_{\mu\ \ \b}^{\ \nu}G^{\a\b}\,.
\end{align}

\paragraph{Expansion of $\sG_6[\a]$.} 
In order to  get explicit results, let us consider the expansion of $\sG_6[\alpha]$ up to $\cO(\alpha^4)$. We learn from the expansion of the characters \eqref{eq:character-K-scalar-UV-local} and \eqref{eq:Q10-expansion} that
\begin{align}
    \sG_6[\alpha]
    &= -\frac{1}{2}\a^2\delta \cF_{\ib\jb}\delta \cF^{\ib\jb}-\frac{\a^4}{24}\varrho_{\cK}[4]+\cO(\a^6)\,, \nn\\
    \varrho_{\cK}[4] &:=  4\delta \cF^{\ib\jb}\delta \cF_{\jb\kb}\delta \cF^{\kb\mb}\delta \cF_{\mb\ib}-3\delta \cF^{\ib\jb}\delta \cF_{\ib\jb}\delta \cF^{\mb\nb}\delta\cF_{\mb\nb}\,,
\end{align}
\normalsize
where 
all higher-order contributions have even power in $\a$ due to maximal supersymmetry. 
Then, 
\normalsize
\begin{align}
    \Gamma_{\rm 1-loop}^{\textnormal{grav}}=&-\frac{\im}{8(4\pi)^2}\int \frac{\sqrt{G}}{\rho^2}\int d\alpha\Tr_{\hs\otimes \mg}\Bigg[e^{-\im\alpha\,(\Box_6 + \Delta_{\tJ}^2) }\sH_{\textnormal{grav}}\,\delta\cF^{\ib\jb}\delta\cF_{\ib\jb}\Bigg]\nn\\
    &-\frac{\im}{96(4\pi)^2}\int \frac{\sqrt{G}}{\rho^2}\int d\alpha\,\alpha^2\Tr_{\hs\otimes \mg}\Bigg[e^{-\im\alpha\,(\Box_6+ \Delta_{\tJ}^2)}\sH_{\textnormal{grav}}\,\varrho_{\cK}[4]\Bigg]\nn\\
    &+\cO(\a^4\delta \cF_{\ib\jb}^6)\,.
\end{align}
\normalsize
Upon integrating out the Schwinger parameter $\a$, we obtain\footnote{Recall that the $\im\epsilon$ prescription is assumed to be encoded in $\Box_6+\Delta_{\tJ}^2$ or $\a\to\a-i\varepsilon$.}
\begin{align}
\label{grav-action-leading-expand}
    \Gamma_{\rm 1-loop}^{\textnormal{grav}}=&-\frac{1}{8(4\pi)^2}\int \frac{\sqrt{G}}{\rho^2}\Tr_{\hs\otimes\mg}\Big(\frac{\sH_{\textnormal{grav}}}{\Box_6+ \Delta_{\tJ}^2}\delta\cF_{\ib\jb}\delta\cF^{\ib\jb}\Big)\nn\\
    &+\frac{1}{48(4\pi)^2}\int \frac{\sqrt{G}}{\rho^2}\Tr_{\hs\otimes\mg}\Big(\frac{\sH_{\textnormal{grav}}\,}{(\Box_6+ \Delta_{\tJ}^2)^3}\,\varrho_{\cK}[4]\Big)\nn\\
    &+\cO(\a^4\delta \cF_{\ib\jb}^6)\,.
    \end{align}
Below, we will show that the contributions associated to higher-order in $\delta \cF_{\ib\jb}$ are sub-leading. Therefore, the quantum behavior of $\Gamma_{\rm 1-loop}^{\textnormal{grav}}$ is mainly described by the leading contribution associated with the $\sH_{\textnormal{grav}}\delta \cF_{\ib\jb}^2$ term.

\subsection{Effective Newton coupling constant}\label{sec:G-Newton}
The $\sH_{\textnormal{grav}}$ term quadratic in the torsion of the Weitzenböck connection can be related to the Ricci scalar $\cR$ of the Levi-Civita connection $\nabla$ associated with the effective metric $G_{\mu\nu}$, via 
\begin{equation}
    \mathcal{R}=-\frac{1}{2}\sH_{\textnormal{grav}}
    -\frac{1}{2}\widetilde{T}_\mu\widetilde{T}_\nu G^{\mu\nu}+2\rho^{-2}G^{\mu\nu}\partial_{\mu}\rho\partial_{\nu}\rho-2\nabla^\mu
    (\rho^{-1}\partial_{\mu}\rho)
\end{equation}
where $\widetilde{T}_\mu$ denotes the $G$-Hodge dual to the totally antisymmetric part of the torsion.
Thus, the Einstein Hilbert action is part of the one-loop effective action above, i.e. 
\begin{equation}
    \Gamma^{\textnormal{grav}}_{\rm 1-loop}\supset S_{\rm EH}:=\int\limits_{\cM^{1,3}}\frac{
    \sqrt{G}}{16\pi G_N}\mathcal{R}\,.
\end{equation}
This leads to 
the identification of $G_N$ with
\begin{align}
\frac{1}{G_N}&\equiv
\frac{1}{4\pi\rho^2}\Tr_{\hs\otimes\mg}\Big(\frac{\delta\cF_{\ib\jb}\delta\cF^{\ib\jb}}{\Box_6+ \Delta_{\tJ}^2}\Big)+\cO(\delta \cF_{\ib\jb}^4)
\end{align}
to leading order in $\delta \cF_{\ib\jb}$. An explicit estimation of the effective Newton coupling constant can be done as follows. 

\paragraph{Evaluating $\Tr_{\mg}$.} 
The trace $\Tr_{\mg}$ can be evaluated using the result of \cite{Steinacker:2024huv}. Suppose that $\cK$ is large and irreducible; then the  trace over $\mg:=\End(\cH_{\cK})$ can be written as 
\begin{align}\label{eq:geo-trace}
    \Tr_{\mg}\Big(\frac{1}{(\Box_6+\Delta_{\tJ}^2)^n}\cO^j\Big)=\int_{\cK\times \cK}\frac{\Omega_{\cK}\times \Omega_{\cK}}{(2\pi)^{|\cK|}}\frac{\big[\cO(x)-\cO(y)\big]^j}{\big[m^2_{\cK}(x-y)^2+\Delta_{\cK}^2+\Delta_{\tJ}^2\big]^n}\,.
\end{align}
Here, we have evaluated the operator $\cO(\cF_{\ib\jb})$ and $\Box_6$ on open string modes $\left|^x_{y} \right)\equiv |x)( y|$ that end on two different point $x,y\in \cK$. 
It is worth noting that the embedding coordinates $x,y$ are dimensionless and normalized as $x^2=\cO(1)=y^2$.  For simplicity, we will replace the integral over $\cK$ with an integral of $x$ over 
a unit sphere, after having rescaled the  radius with the scale factor (i.e. KK scale) $m_\cK$. 
Furthermore, $\Omega_{\cK}$ is the symplectic volume form of $\cK$ and $|\cK|=\dim(\cK)$.

With the above setting, we can obtain the leading  contributions to the Newton constant
explicitly as
\begin{equation}
    \frac{1}{G_N}\approx \frac{1}{4\pi \rho^2}\Big(\vartheta_\cK[2]-\frac{1}{6}\vartheta_\cK[4]\Big)+\cO\Big(\vartheta_{\cK}[n\geq 6]\Big)\,,
\end{equation}
where
\begin{subequations}
    \begin{align}
    \vartheta_{\cK}[2]:&=\Tr_{\mg}\Big(\frac{1}{\Box_6+\Delta_\tJ^2}\delta \cF_{\ib\jb}\delta \cF^{\ib\jb}\Big)\,,\\
    \vartheta_{\cK}[4]:&=\Tr_{\mg}\Bigg[\frac{1}{(\Box_6+\Delta_\tJ^2)^3}\Big(4\delta \cF^{\ib\jb}\delta \cF_{\jb\kb}\delta \cF^{\kb\mb}\delta \cF_{\mb\ib}-3\delta \cF^{\ib\jb}\delta \cF_{\ib\jb}\delta \cF^{\mb\nb}\delta\cF_{\mb\nb}\Big)\Bigg]\,.
\end{align}
\end{subequations}
To compare these two contributions, we shall evaluate them explicitly as follows:

\begin{itemize}
    \item[$\cdot$] {\bf Evaluating $\vartheta_{\cK}[2]$.} Using the geometric trace formula \eqref{eq:geo-trace}, 
    \begin{align}
    \Tr_{\mg}\Big(\frac{\delta\cF^{\ib\jb}\delta\cF_{\ib\jb}}{\Box_6+\Delta_\tJ^2}\Big)=\int_{\cK\times \cK}\frac{\Omega_{\cK}\times \Omega_{\cK}}{(2\pi)^{|\cK|}}\frac{(\cF(x)-\cF(y))^2}{m_{\cK}^2(x-y)^2+\Delta_{\cK}^2+\Delta_\tJ^2}\,.
\end{align}
Since $\cF_{\ib\jb}\sim \cO(\Delta_{\cK}^2)$ and $m_{\cK}^2=d_{\cK}^{2/|\cK|}\Delta^2_{\cK}$, the above can be estimated as
\begin{align}\label{eq:trace-over-single-K-1}
    \Tr_{\mg}\Big(\frac{\delta\cF^{\ib\jb}\delta\cF_{\ib\jb}}{\Box_6+\Delta_\tJ^2}\Big)\sim d_{\cK}^2\Delta^2_{\mathcal{K}}\int_{\cK\times \cK} d^{|\mathcal{K}|}x\,d^{|\cK|}y\frac{1}{{\frac{m^2_{\cK}}{\Delta^2_{\mathcal{K}}}(x-y)^2+1+\frac{\Delta_{\tJ}^2}{\Delta_{\cK}^2}}}\,,
\end{align}
where we have used the Bohr-Sommerfeld quantization condition \eqref{eq:Bohr-Q} to carry out the factor $d_{\cK}^2$ in front. Upon making a change of variables $r^{\ib}=x^{\ib}-y^{\ib}$ and $\tilde r^{\ib}=x^{\ib}+y^{\ib}$ where $\ib=1,\ldots,|\cK|$, 
\begin{align}
    \vartheta_{\cK}[2]\sim d_{\cK}^2\Delta_{\cK}^2\int_{\cK\times \cK}d^{|\cK|}r \,d^{|\cK|}\tilde r\frac{1}{d^{2/|\cK|}_{\cK}r^2+1+\frac{\Delta_{\tJ}^2}{\Delta_{\cK}^2}}\,.
\end{align}
The integral over $\tilde r$ results in a number of order $\cO(1)$, and the integral over $r$ gives
\small
\begin{subequations}
    \begin{align}
        |\cK|=2&: &\vartheta_{\cK}[2]&\sim d_{\cK}\Delta_{\cK}^2\log\Big(1+\frac{d_{\cK}\Delta_{\cK}^2}{\Delta_{\tJ}^2+\Delta_{\cK}^2}\Big)\,,\\
        |\cK|=4&: &\vartheta_{\cK}[2]&\sim d_{\cK}\Delta_{\cK}^2\Big[\sqrt{d_{\cK}}+\Big(1+\frac{\Delta_{\tJ}^2}{\Delta_{\cK}^2}\Big)\log\Big(\frac{\Delta_{\tJ}^2+\Delta_{\cK}^2}{\Delta_{\tJ}^2+\sqrt{d_{\cK}}\Delta_{\cK}^2}\Big)\Big]\,,\\
        |\cK|=6&: &\vartheta_{\cK}[2]&\sim d_{\cK}\Big[d_{\cK}^{1/3}\Big(d^{1/3}\Delta_{\cK}^2-2\Delta_{\tJ}^2\Big)+\frac{(\Delta_{\tJ}^2+\Delta_{\cK}^2)^2}{\Delta_{\cK}^2}\log\Big(\frac{\Delta_{\tJ}^2+d^{1/3}_{\cK}\Delta_{\cK}^2}{\Delta_{\tJ}^2+\Delta_{\cK}^2}\Big)\Big]\,.
    \end{align}
\end{subequations}
\normalsize
As alluded to in the above, since $\Delta_{\tJ}$ is an IR scale and $\Delta_{\cK}$ is a UV scale, it is reasonable to assume that $\Delta_{\tJ}\ll \Delta_{\cK}$. Under this assumption,
\begin{subequations}
    \begin{align}
        |\cK|=2&: &\vartheta_{\cK}[2]&\sim d_{\cK}\Delta_{\cK}^2\log\Big(1+d_{\cK}\Big)\,,\\
        |\cK|=4&: &\vartheta_{\cK}[2]&\sim d_{\cK}\Delta_{\cK}^2\frac{\sqrt{d_{\cK}} - \log(1 + \sqrt{d_{\cK}})}{2}\,,\\
        |\cK|=6&: &\vartheta_{\cK}[2]&\sim d_{\cK}\Delta_{\cK}^2 \frac{ d_{\cK}^{2/3}-2 d_{\cK}^{1/3}  + 2 \log(1 + d_{\cK}^{1/3})}{4 }\,;
    \end{align}
\end{subequations}
we recall that $|\cK|$ is the dimension of $\cK$. 
    \item[$\cdot$] {\bf Evaluating $\vartheta_{\cK}[4]$.} The evaluation of $\vartheta_{\cK}[4]$ can be done similarly. For large $d_{\cK}$ and $\Delta_{\tJ}\ll \Delta_{\cK}$, we have
    \begin{subequations}
    \begin{align}
        |\cK|=2&: &\vartheta_{\cK}[4]&\sim d_{\cK}\Delta_{\cK}^2\,,\\
        |\cK|=4&: &\vartheta_{\cK}[4]&\sim d_{\cK}\Delta_{\cK}^2\,,\\
        |\cK|=6&: &\vartheta_{\cK}[4]&\sim d_{\cK}\Delta_{\cK}^2\log(d_{\cK}) \,.
    \end{align}
\end{subequations}
\end{itemize}
Observe that since $d_{\cK}\gg 1$ for large $\cK$, 
\begin{align}
    \frac{\vartheta_{\cK}[2]}{\vartheta_{\cK}[4]}\sim \frac{1}{d_{\cK}}\ll 1\,.
\end{align}
We conclude that $\vartheta_{\cK}[2]$ gives  the leading contribution in the one-loop effective action for the (higher-spin extended) gravity on $\cM^{1,3}$ induced by the IKKT matrix model, in the presence of ``large" extra dimensions $\cK$, i.e. large $d_\cK$. 
We will see in Section \ref{sec:higher-derivatives}
that all higher-order (higher curvature) contributions are suppressed similarly.

\paragraph{Evaluating $\Tr_{\hs}$.} Now consider the trace over the $\hs$ modes in
\begin{align}\label{eq:Gamma-GR}
    \Gamma_{\rm 1-loop}^{\textnormal{grav}}\approx&-\frac{1}{4(4\pi)^2}\int \frac{\sqrt{G}}{\rho^2}\Tr_{\hs}\Bigg[\sH_{\textnormal{grav}}\Bigg(\vartheta_{\cK}[2]-\frac{\vartheta_{\cK}[4]}{6}+\ldots\Bigg)\Bigg]\,.
\end{align}
Note that we can effectively write the Poisson brackets as
\begin{subequations}\label{eq:Poisson-oblivion}
    \begin{align}
 \{x^\nu,\sum_{s,m}\phi_{sm}(x)Y^{sm}\} &\approx \sum_{s,m}\{x^\nu,\phi_{sm}(x)\}Y^{sm}\,,\\ 
 \{\ttb^{\dot\nu},\sum_{s,m}\phi_{sm}(x)Y^{sm}\} &\approx \sum_{s,m}\{\ttb^{\dot\nu},\phi_{sm}(x)\}Y^{sm} \, ,
\end{align}
\end{subequations}
since other contributions such as $\{x^\nu,Y^{sm}\}\phi_{sm}$ would have fewer space-time derivatives and hence do not contribute to gravity.
As a consequence, we can simply pull out all $\hs$ factors $Y^{sm}$, and  the trace over $\hs$-modes can be done in the same way as in  \cite{Steinacker:2024huv}. 
It simply results in some factors $\tJ$  or $\tJ^2$ as follows\footnote{Here, the scenario is that the number of $\hs$ modes, whose mass terms $m_s^2$ can be neglected in the loops in the IR regime, are truncated at $\tJ$. This holds in the semi-classical regime of the present background $\cM_\tJ^{1,3}$, where $\tJ$ marks the transition to the "deep quantum`` regime of the geometry. However, the truncation in spin is not strict and needs to be understood in more detail.}.
Let us normalize the $Y$'s as 
\begin{align}
    \tr_{\C^{\tJ}}\Big(Y^{\ell_1m_1}Y^{\ell_2m_2}\Big)=\tJ\, \delta^{\ell_1,\ell_2}\delta^{m_1+m_2,0}\,.
\end{align}
Then, the trace over $\hs$-modes result in
\begin{subequations}
    \begin{align}
        |\cK|=2&: &\Tr_{\hs}\Big(\sH_{\textnormal{grav}}^{\hs}\vartheta_{\cK}[2]\Big)&\sim \tJ^2d_{\cK}\Delta_{\cK}^2\log\Big(1+\frac{d_{\cK}\Delta_{\cK}^2}{\Delta_{\tJ}^2+\Delta_{\cK}^2}\Big)\,,\\
        |\cK|=4&: &\Tr_{\hs}\Big(\sH_{\textnormal{grav}}^{\hs}\vartheta_{\cK}[2]\Big)&\sim \tJ^2\,d_{\cK}\Delta_{\cK}^2\Big[\sqrt{d_{\cK}}+\Big(1+\frac{\Delta_{\tJ}^2}{\Delta_{\cK}^2}\Big)\log\Big(\frac{\Delta_{\tJ}^2+\Delta_{\cK}^2}{\Delta_{\tJ}^2+\sqrt{d_{\cK}}\Delta_{\cK}^2}\Big)\Big]\,
    \end{align}
\end{subequations}
given that $\tr_{\C^\tJ}=\tJ$.\footnote{As a reminder, we do trace twice over all $\hs$-modes.} For $\Delta_{\cK}\gg\Delta_{\tJ}$, the above simplifies to
\begin{subequations}
    \begin{align}
        |\cK|=2&: &\Tr_{\hs}\Big(\sH_{\textnormal{grav}}\vartheta_{\cK}[2]\Big)&\sim d_{\cK}^2 \Delta_{\cK}^2 \tJ^{2}\sH_{\textnormal{grav}}\,,\\
        |\cK|=4&: &\Tr_{\hs}\Big(\sH_{\textnormal{grav}}\vartheta_{\cK}[2]\Big)&\sim d_{\cK}^{3/2} \Delta^2_{\cK} \tJ^2\,\sH_{\textnormal{grav}}\,.
    \end{align}
\end{subequations}


\paragraph{Effective Newton coupling.} Combining the above results, we can read off the effective Newton constant explicitly. Focusing on $|\cK|=4$ case, we obtain
\begin{align}
\label{eq:G-Newton-constant}
    \frac{1}{G_N}:\approx\frac{\tJ^2\,d_{\cK}^{3/2}}{\pi}\frac{\Delta_{\cK}^2}{\rho^2}\,.
\end{align}
Here, the choices $|\cK|=2,4$ are natural from the point of view of stabilizing $\cK$ and having non-trivial vacuum as explained below. 
In particular, we observe that the effective Planck scale $\frac{1}{G_N}$ also evolves as $\frac{\Delta_{\cK}^2}{\rho^{2}}$ during 
  the cosmic evolution similarly with 
  all other effective KK scales, cf. \eqref{evolution-KK-masses}. This means that the resulting local physics -- expressed in terms of these scales -- is time-independent, consistent with observation. 






\section{Stabilization of \texorpdfstring{$\cK$}{K} and dynamics of \texorpdfstring{$\Delta_{\cK}$}{DK}}\label{sec:stabilize}
The fuzzy extra dimensions $\cK$ naturally induce a Kaluza Klein scale $m_{\cK}$, which breaks the conformal symmetry of the $\cN=4$ SYM sector in \hsikkt,  leading to plausible low-energy physics.\footnote{Note that $\cM^{1,3} \times\cK$ should be thought of as embedded in the uncompactified target space $\R^{1,9}$. Therefore, we do not face the vast landscape problem as in string theory.} This is possible because the effective potential associated to the $SO(1,9)$ character $\sQ_{10}$ cf. \eqref{eq:Q10} has a global minimum, unlike in the conventional $\cN=4$ SYM. Here, we attempt to estimate the dynamics of the non-commutativity scale $\Delta_{\cK}$ as a function of the dilation $\rho$. 

\subsection{Revisiting the classical action for \texorpdfstring{$\cK$}{K}}
Let us recall the semi-classical action for the internal flux $\cF_{\ib\jb}$ on the background brane $\cM^{1,3} \times \cK$. Here we focus on the $\cK$-sector, which is governed by the following classical action
\begin{align}
    S_0[\cF_{\ib\jb}]=\frac{1}{g^2}\tr\Big([t_{\ib},t_{\jb}][t^{\ib},t^{\jb}]\Big)=-\frac{1}{g^2}\tr\Big(\cF_{\ib\jb}\cF^{\ib\jb}\Big)\,,\qquad \cF_{\ib\jb}:=-\im [t_{\ib},t_{\jb}]\,.
\end{align}
Using the trace in the semi-classical regime, cf. \eqref{eq:classical-trace}, we obtain
\begin{align}
    S_0[\cF_{\ib\jb}]=-\frac{1}{\ell_p^4\,g^2}\int \frac{\sqrt{G}}{\rho^2}\tr_{\C^{\tJ}\otimes\cH_{\cK}}(\cF_{\ib\jb}\cF^{\ib\jb})=-\frac{\tJ}{\ell_p^4\,g^2}\int \frac{\sqrt{G}}{\rho^2}\tr_{\cH_{\cK}}(\cF_{\ib\jb}\cF^{\ib\jb}).
\end{align}
Note that since $[g]=mass^2$, it is natural to define the following \emph{dimensionless} coupling
\begin{align}
    \gp:=\ell_p^2\,g
\end{align}
which governs the classical action $S_0[\cF_{\ib\jb}]$.\footnote{Note that we could also incoporate $\tJ$ in the definition of $\gp$. However, the present form is more natural for relating the classical potential with the one loop effective action.} This  coupling can be related to the classical Yang-Mills coupling  for nonabelian gauge fields in the present model via \cite{Steinacker:2024huv}
\begin{align}
\label{YM-coupling}
   \frac 1{g_{\rm YM}^2} =  \frac{\tJ}{\gp^2} \rho^2
   \,,\qquad\text{or}\qquad g_{\rm YM}^2=\frac{\gp^2}{\tJ\,\rho^{2} }\,.
\end{align}
Thus, we may also write $S_0[\cF_{\ib\jb}]$ as
\begin{align}
     S_0[\cF_{\ib\jb}]=-\int \frac{\sqrt{G}}{\rho^4}\frac{1}{g^2_{\rm YM}}\tr_{\cH_{\cK}}(\cF_{\ib\jb}\cF^{\ib\jb})\,.
\end{align}

\subsection{Effective potential for \texorpdfstring{$m_\cK$}{mk}}
Let us now discuss the dynamics of $m_{\cK}$. Of course, the present approach only makes sense if $\cK$ is stable. Thus, let us recall the result of \cite{Steinacker:2024huv,Steinacker:2023myp} where  the various contributions to the effective potential for the radius $m_\cK$ (or equivalently $\Delta_\cK$) were computed, and elaborate how the (non-trivial) minimum $m_\cK$ depends on $\rho$ and the other parameters of the background.

For simplicity, we restrict ourselves to the case of a single, large $d_\cK$ brane $\cK$. Then, the effective potential $V_{\textnormal{eff}}$ comprises the following pieces
\begin{align}\label{eq:Veff}
    V_{\textnormal{eff}}=V_0+V^{\rm 1-loop}_{\cK}+V_{S^2_{\tJ}-\cK}+V_{S^2_{\tJ}}
    +V_{\textnormal{grav}}
\end{align}
where $V_0$ is the classical potential, $V^{\rm 1-loop}_{\cK}$ is the potential coming from the one-loop effective action 
for $\cK$, $V_{S^2_{\tJ}-\cK}$ is the potential between $S^2_{\tJ}$ and $\cK$, and $V_{S^2_{\tJ}}$ is the potential associated to the flux on $S^2_{\tJ}$. We also included the potential $V_{\textnormal{grav}}$ associated with the gravitational background, which is based on $\Gamma_{\rm 1-loop}^{\textnormal{grav}}$ cf. \eqref{eq:Gamma-GR}.

\paragraph{$V_0$ potential.} By definition $S_0[\cF_{\ib\jb}]=-\int \sqrt{G}\,V_0$, with
\begin{align}\label{eq:V0}
    V_0=\frac{\tJ}{g^2_{\rm YM}\rho^4}\tr_{\cH_{\cK}}\Big(\cF_{\ib\jb}\cF^{\ib\jb}\Big)\approx \frac{\tJ\Delta_{\cK}^4 d_\cK}{g^2_{\rm YM}\,\rho^4}>0\,.
\end{align}

\paragraph{$V^{\rm 1-loop}_{\cK}$ potential.} This potential is computed from the first non-trivial contribution $\sX_6$. Here, we simply recall the result of \cite{Steinacker:2024huv}
\small
\begin{align}\label{eq:potential-for-K}
    V^{\rm 1-loop}_{\cK}
    &=-\frac{\tJ^2\,\pi^2}{ \rho^4}
 \int\limits_{\cK_x\times \cK_y}\! \frac{\Omega_x\times\Omega_y}{(2\pi)^{\frac{|\cK_x|+|\cK_y|}{2}}}\,\frac{4\Tr(\cF(x)-\cF(y))^4 -(\Tr(\cF(x)-\cF(y))^2)^2}{(m_{\cK}^2(x-y)^2+\Delta_{\cK}^2)^2}\,.
\end{align}
\normalsize
In the setting where $\cK$ is a single large brane, we have
\begin{subequations}
    \begin{align}
         |\cK|=2&: &V^{\rm 1-loop}_{\cK}&\sim -\tJ^2\Delta_{\cK}^4\,\frac{d_{\cK}}{\rho^4}\,,\\
        |\cK|=4&: &V^{\rm 1-loop}_{\cK}&\sim -\tJ^2\Delta_{\cK}^4\, \frac{d_{\cK}\log(d_{\cK})}{\rho^4}\,,\\
        |\cK|=6&: &V^{\rm 1-loop}_{\cK}&\sim -\tJ^2\Delta^4_{\cK}\, \frac{d^{4/3}_{\cK}}{\rho^4} \,.
    \end{align}
\end{subequations}
Essentially, this potential gives a negative correction to $V_0$, as both are proportional to $\Delta^4_{\cK}$. 
The resulting stability requirements  will be discussed in Section \ref{sec:dynamics-Delta-K}.

\paragraph{$V_{S^2_{\tJ}-\cK}$ potential.} This is the crucial contribution in $V_{\textnormal{eff}}$ that leads to bound states and allows $\cK$ to introduce a scale and break conformal symmetry of the $\cN=4$ SYM sector. As the notation suggests, $V_{S^2_{\tJ}-\cK}$ is obtained from interactions between the background $U(1)$ fluxes $\cF^{\mu\nu}$ on $S^2_{\tJ}$, and fluxes on $\cK$. 
 This contribution arises from restricting $\sQ_{10}$ to the $\delta\cF^{\mu\nu}\delta\cF_{\mu\nu}\delta\cF_{\ib\jb}\cF^{\ib\jb}$ term. 
The corresponding potential reads
(cf. (7.22) in \cite{Steinacker:2024huv})
\small
\begin{align}\label{eq:potential-S2-K}
V_{S^2_{\tJ}-\cK}^{\rm 1-loop} &= -\rho^{-4} \Tr_{\hs}\Big(\delta\cF^{0i}\delta \cF^0{}_{i}\Big)
\int\limits_{\cK \times \cK} 
\frac{\Omega_x \times \Omega_y}{(2\pi)^{|\cK|}}\,
\frac{(\cF(x)-\cF(y))^2}{[m_\cK^2(x-y)^2 + \Delta_\cK^2 + \Delta_{\tJ}^2]^2}\, .
\end{align}
\normalsize
Note that we have used $ \delta\cF^{0i}\delta \cF_{0i} = - \delta\cF^{0i}\delta \cF^0_{\ i}$. Then, since $\cF_{\ib\jb}\sim \cO(\Delta_{\cK}^2)$, we get 
\begin{align}
V_{S^2_{\tJ}-\cK}^{\rm 1-loop} \approx \frac{\Delta_{\cK}^4}{\rho^{4}} I_{S^2_{\tJ}}\times I_{\cK} 
\end{align}
where $I_{S^2_{\tJ}}$ and $I_{\cK}$ are two contributions that are computed below. 
First we need the flux on $S^2_\tJ$, which is given by
\begin{align}
     \cF^{0i}=-\frac{1}{\ell_p^2R^2}\theta^{0i}=-\frac{1}{R}t^i\,,\qquad i=1,2,3\,
\end{align}
recalling that $\theta^{0i}
\sim \ell_p^2R t^i$. 
Then the first factor is
\begin{align}\label{eq:Int-S2-1}
    I_{S^2_{\tJ}}\equiv\Tr_{\hs}\Big([\cF^{0i},[\cF^{0}{}_i,-]]\Big)=\frac  1{R^2}\Tr_{\hs}\Big([t^i,[t_i,-]]\Big)\approx \frac{\tJ^4}{R^4}\sim \frac{1}{\ell_p^4}\,,
\end{align}
where we have used
\begin{align}
    \Tr_{\hs}\Big([t^i,[t_i,-]]\Big)=\frac{1}{R^2}\sum_{s=0}^\tJ(2s+1)s(s+1)\, \sim \frac{\tJ^4}{R^2} \,.
\end{align}
The second factor $I_{\cK}$
\small
\begin{align}\label{eq:IK}
   I_\cK\equiv -\int\limits_{\cK \times \cK} 
\frac{\Omega_x\times \Omega_y}{(2\pi)^{|\cK|}}\,
\frac{1}{[m_\cK^2(x-y)^2 + \Delta_\cK^2 + \Delta_{\tJ}^2]^2}  
\end{align}
\normalsize
can be estimated as 
\begin{subequations}
    \begin{align}
        |\cK|=2&: &I_{\cK}&\sim -\frac{d_{\cK}}{ \Delta_{\cK}^2 (\Delta_{\tJ}^2 + \Delta_{\cK}^2)}\,,\\
        |\cK|=4&: &I_{\cK}&\sim +\frac{d_{\cK}}{\Delta_{\cK}^4}\Bigg[\frac{d_{\cK}^{1/2}\Delta_{\cK}^2}{\Delta_{\tJ}^2+d_{\cK}^{1/2}\Delta_{\cK}^2}+\log\Big(\frac{\Delta_{\tJ}^2+\Delta_{\cK}^2}{\Delta_{\tJ}^2+d_{\cK}^{1/2}\Delta_{\cK}^2}\Big)\Bigg]\,,\\
        |\cK|=6&: &I_{\cK}&\sim -\frac{d^{4/3}_{\cK}}{2 \Delta_{\cK}^4} \,,
    \end{align}
\end{subequations}
assuming $d_{\cK}$ is large. Altogether, we obtain
\begin{subequations}
    \begin{align}
        |\cK|=2&: &V_{S^2_{\tJ}-\cK}&\sim -\frac{1}{\ell_p^4}\frac{d_{\cK}\Delta_{\cK}^2 }{\rho^{4} (\Delta_{\tJ}^2 + \Delta_{\cK}^2)}\,,\\
        |\cK|=4&: &V_{S^2_{\tJ}-\cK}&\sim + \frac{1}{\ell_p^4}\frac{d_{\cK}}{\rho^{4}}\Bigg[\frac{d_{\cK}^{1/2}\Delta_{\cK}^2}{\Delta_{\tJ}^2+d_{\cK}^{1/2}\Delta_{\cK}^2}+\log\Big(\frac{\Delta_{\tJ}^2+\Delta_{\cK}^2}{\Delta_{\tJ}^2+d_{\cK}^{1/2}\Delta_{\cK}^2}\Big)\Bigg]\,,\label{eq:good-VS2-K}\\
        |\cK|=6&: &V_{S^2_{\tJ}-\cK}&\sim -\frac{1}{\ell_p^4}\frac{d^{4/3}_{\cK}}{\rho^{4}} \,.
    \end{align}
\end{subequations}
Already at this stage, we can conclude that $|\cK|$ must be 2 or 4 in order to have a non-trivial minimum of the potential (as a function of $\Delta_\cK$) in this approximation. Then $V_{S^2_{\tJ}-\cK}$ vanishes for $\Delta_\cK = 0$, and the leading term in the Taylor expansion in $\Delta_{\cK}$ around the origin is negative, i.e.
\begin{subequations}
    \begin{align}
    |\cK|=2&: &V_{S^2_{\tJ}-\cK}^{\rm 1-loop} &\sim - \frac{1}{\ell_p^4}\frac{d_{\cK}}{\rho^4}\frac{\Delta_{\cK}^2 }{ \Delta_{\tJ}^2}\,,\\
    |\cK|=4&: &V_{S^2_{\tJ}-\cK}^{\rm 1-loop} &\sim - \frac{1}{\ell_p^4}\frac{d_{\cK} }{\rho^{4}} \frac{\Delta_\cK^4}{\Delta_\tJ^4}\,.
    \label{V-S2-K-K4}
\end{align}
\end{subequations}
This indicates that $|\cK|=2,4$ should be favored when extracting interesting low energy physics from the IKKT matrix model on the proposed $\cM^{1,3}_{\tJ}\times \cK$ background. 


\paragraph{$V_{S^2_{\tJ}}$ potential.} This potential is associated with the contribution  $\sQ_{10}\Big|_{\delta\cF_{\mu\nu}^4}=:\Sigma_4(\delta \cF_{\mu\nu})$, which can be evaluated as \cite{Steinacker:2024huv}
\begin{align}
    \Sigma_4(\delta \cF_{\mu\nu})
    &=4\delta \cF^{\mu\nu}\delta \cF_{\nu\rho}\delta \cF^{\rho\sigma}\delta \cF_{\sigma\mu}
   - (\delta \cF^{\mu\nu}\delta \cF_{\mu\nu})^2 \nn\\
&= 4\big(2 (\delta \cF^{0j}\delta \cF_{j0})^2
    + 4\delta \cF^{0j}\delta \cF_{jk}\delta \cF^{kl}\delta \cF_{l 0} 
+  \delta \cF^{ik}\delta \cF_{kl}\delta \cF^{ln}\delta \cF_{n i}\big) - (\delta \cF^{\mu\nu}\delta \cF_{\mu\nu})^2 \nn\\
&\approx 4\big(2 (\delta \cF^{0j}\delta \cF_{j0})^2\big) - (2\delta \cF^{0i}\delta \cF_{0i})^2 \nn\\
&\approx 4(\delta \cF^{0j}\delta \cF_{j0})^2 >0\,
\end{align}
dropping the purely space-like contributions which are subleading. Then, the potential $V_{S^2_{\tJ}}$ associated with $\Sigma_4$ reads
\begin{align}
V_{S^2_{\tJ}} &= -\frac{4}{\rho^{4}}  \Tr_{\hs}\Big([\cF^{0i},[\cF^0{}_{i},[\cF^{0j},[\cF^0{}_{j},-]]]]\Big)
\int\limits_{\cK \times \cK} 
\frac{\Omega_x \times \Omega_y}{(2\pi)^{|\cK|}}\,
\frac{1}{[m_\cK^2(x-y)^2 + \Delta_\cK^2 + \Delta_{\tJ}^2]^2}  \,.
\end{align}
\normalsize
The trace $\Tr_{\hs}(\delta\cF_{0i}^4)$ is evaluated to
\begin{align}
   \Tr_{\hs}\Big([\cF^{0i},[\cF^0{}_{i},[\cF^{0j},[\cF^0{}_{j},-]]]]\Big)=\frac{1}{R^4}\Tr_{\hs}\Big([t^i,[t_i,[t^j,[t_j,-]]]]\Big)\approx \frac{\tJ^6}{R^8}\sim \frac{1}{\ell_p^6R^2}\,.
\end{align}
As discussed above, $|\cK|=2,4$ is favored 
for interesting physics to emerge. Focusing on these cases, we have the following ratio between $V_{S^2_{\tJ}}$ and $V_{S^2_{\tJ}-\cK}$
\begin{align}\label{eq:ratio-VS2-VSK}
    \frac{V_{S^2_{\tJ}}}{V_{S^2_{\tJ}-\cK}}\sim\frac{1}{\ell_p^2R^2\Delta_{\cK}^4}\sim \frac{\tJ^2}{R^4\Delta_{\cK}^4}\,.
\end{align}

\paragraph{Gravitational contribution.} Recall that the induced gravitational action is given by \cite{Steinacker:2023myp}
\begin{align}
    \Gamma_{\rm 1-loop}^{\textnormal{grav}}=
     -\int\limits_\cM \frac{\sqrt{G}}{32\pi G_N} \sH^{\textnormal{grav}}\,,\qquad  \frac{1}{G_N}:\approx\frac{\tJ^2\,d_{\cK}^{3/2}}{\pi}\frac{\Delta_{\cK}^2}{\rho^2}\,,
\end{align}
where we have computed the effective Newton coupling constant above cf. \eqref{eq:G-Newton-constant}. Hence, we can define the gravitational potential as 
\begin{align}
    V_{\textnormal{grav}}= 
    \frac{1}{32\pi\,G_N} \sH^{\textnormal{grav}}\,.
\end{align}
Its background value can be obtained by first evaluating \cite{Battista:2023glw}
\begin{align}\label{eq:BG-torsion-square}
    \cT^{\dot\mu\dot\nu\alpha}\cT_{\dot\mu\dot\nu}{}^{\alpha}\gamma_{\a\a}\Big|_{t^{\dot\mu}\rightarrow \ttb^{\dot\mu}}&=\Big(E^{\dot\mu\mu}\p_{\mu}E^{\dot\nu\alpha}-E^{\dot\nu\mu}\p_{\mu}E^{\dot\mu\alpha}\Big)\Big(E_{\dot\mu}{}^{\sigma}\p_{\sigma}E_{\dot\nu}{}^{\alpha}-E_{\dot\nu}{}^{\sigma}E_{\dot\mu}{}^{\alpha}\Big)\gamma_{\a\a}\nn\\
    &\sim \p^{\mu}\sinh(\tau)\p_{\mu}\sinh(\tau)=\p^{\mu}\frac{x_4}{R}\p_{\mu}\frac{x_4}{R}\nn\\    
    &\sim-\frac{1}{R^2}\frac{\cosh^2(\tau)}{\sinh^2(\tau)}<0\,,
\end{align}
where we recall that $x_4=\sqrt{-x^{\mu}x_{\mu}-R^2}$ and $x^{\mu}x_{\mu}=-R^2\cosh^2(\tau)$. Since
\begin{align}
    \cT^{\dot\mu\dot\nu\alpha}\cT_{\dot\mu\dot\nu}{}^{\alpha}\gamma_{\a\a}=\rho^2 \sH^{\textnormal{grav}}\,,
\end{align}
the gravitational potential for the $\cM^{1,3}$ background is evaluated as
\begin{align}
    V_{\textnormal{grav}}\sim \frac{1}{\rho^2}\frac{1}{G_{N}}\cT^{\dot\mu\dot\nu\alpha}\cT_{\dot\mu\dot\nu}{}^{\alpha}\gamma_{\alpha\alpha}\Big|_{t^{\dot\mu}\rightarrow \ttb^{\dot\mu}}\approx-\frac{\tJ^2\Delta_{\cK}^2}{R^2\rho^{4}}d_{\cK}^{3/2}\,.
\end{align}

\subsection{Stabilization of $\cK$} \label{sec:dynamics-Delta-K}

We can now  combine all the above contributions to $V_{\rm eff}(\Delta_\cK)$, and try to establish a non-trivial global minimum. This would also introduce a non-trivial KK scale into the theory.
Focusing on $|\cK|=4$, 
the combined effective potential \eqref{eq:Veff}
is given explicitly by
\begin{align}
    V_{\textnormal{eff}}\approx& \ V_0 + V_\cK^{\rm 1-loop}
   +V_{\textnormal{grav}} +V_{S^2_{\tJ}-\cK}+V_{S^2_{\tJ}}  \nn\\
    &\approx +\Delta_{\cK}^4\Big(\frac{\tJ d_\cK}{g^2_{\rm YM}\,\rho^4}
    -\frac{\tJ^2d_{\cK}\log(d_{\cK})}{\rho^4}\Big) -\frac{\tJ^2\Delta_{\cK}^2}{R^2\rho^{4}}d_{\cK}^{3/2}\nn\\
    &+\frac{\tJ^2\,d_{\cK}}{\ell_p^2R^2\rho^{4}}\Bigg[\frac{\sqrt{d_{\cK}}\Delta_{\cK}^2}{\Delta_{\tJ}^2+\sqrt{d_{\cK}}\Delta_{\cK}^2}-\log\Big(\frac{\Delta_{\tJ}^2+\sqrt{d_{\cK}}\Delta_{\cK}^2}{\Delta_{\tJ}^2+\Delta_{\cK}^2}\Big)\Bigg]\nn\\
    &+\frac{\tJ^2\,d_{\cK}}{\ell_p^4R^4\Delta_{\cK}^4\rho^4}\Bigg[\frac{\sqrt{d_{\cK}}\Delta_{\cK}^2}{\Delta_{\tJ}^2+\sqrt{d_{\cK}}\Delta_{\cK}^2}-\log\Big(\frac{\Delta_{\tJ}^2+\sqrt{d_{\cK}}\Delta_{\cK}^2}{\Delta_{\tJ}^2+\Delta_{\cK}^2}\Big)\Bigg] \,.
\end{align}
Here $V_{\cK}^{\rm 1-loop}$ and $V_0$ are
combined in the first bracket, since both are proportional to $\Delta^4_{\cK}$. 

It was shown in \cite{Steinacker:2024huv} that  this effective potential has a stable non-trivial minnimum at some $\Delta_{\cK}^*>0$,
in the regime where $V_{S^2_{\tJ}}\ll V_{S^2_{\tJ}-\cK}$. 
This is due to the negative contribution \eqref{V-S2-K-K4} near the origin in $\Delta_\cK^4$, 
while $V_0>0$ dominates for large $\Delta_\cK$.
The corresponding non-trivial vacuum clearly breaks the conformal symmetry of the $\cN=4$ SYM sector.

\begin{figure}[ht!]
    \centering
    \includegraphics[scale=0.42]{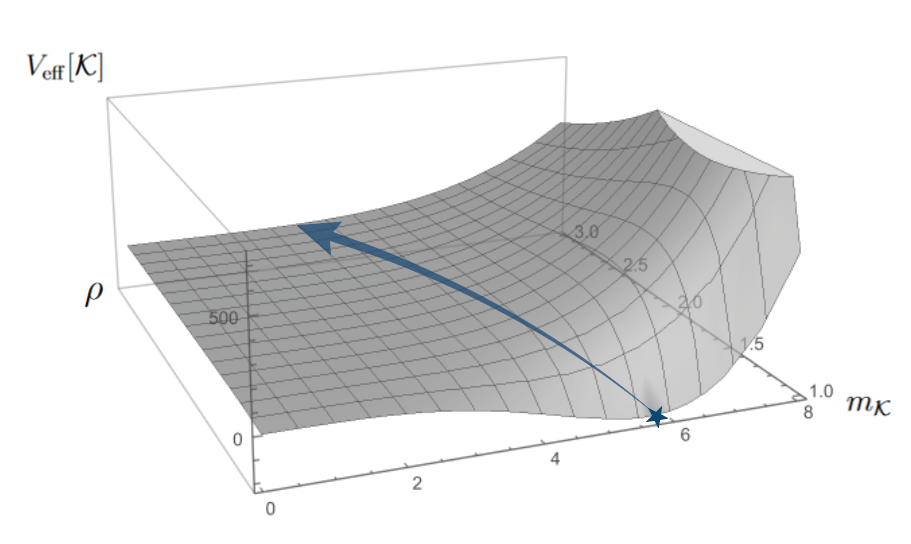}
    \caption{The blue arrow represents the trajectory of the minima [star] of $V_{\rm eff}$ at sufficiently early time. As the universe expands, i.e. as $\rho$ increasing, the effective potential $V_{\rm eff}$ flattens out and the minima approaches zero. }
    \label{fig:Veff-rho}
\end{figure}

Since $V_{\rm eff}$ changes during the cosmic expansion through $\rho$, the minimum of $V_{\rm eff}$ also evolves during the cosmic evolution. In particular, $\Delta_\cK^* \to 0$ at late times, which can be seen in Figure \ref{fig:Veff-rho}. 

\paragraph{Adiabatic approximation for  $\Delta_{\cK}^*$.}



With some extra assumptions,
we can determine the minimum $\Delta_\cK^*$ explicitly in an adiabatic approximation.
To simpify the analysis,
we shall focus on the regime $V_{S^2_{\tJ}}\ll V_{S^2_{\tJ}-\cK}$, which using \eqref{eq:ratio-VS2-VSK} implies
\begin{align}\label{eq:1st-constraint-DK}
    \frac{1}{\ell_pR}\sim \frac{\tJ}{R^2}\ll\Delta_{\cK}^2\,
\end{align}
although this is not strictly necessary.
On the other hand, for large $\Delta_\cK$, $V_0$ should dominate $V_{\cK}^{\rm 1-loop}$ so that $\cK$ can be stable. This amounts to the stability condition\footnote{
This condition is violated at early times, since 
$g_{\rm YM}^{-1}\sim\rho$. However, the kinetic term for $m_\cK$ discussed below would prevent a runaway behavior in this finite regime. Note  that the 1-loop corrected Yang-Mills coupling can be significantly larger (and time-independent) in a sufficiently early period of the cosmic evolution \cite{Steinacker:2024huv}.}
\begin{align}
\label{eq:2nd-constraint-DK}
    g^2_{\rm YM}  \ll\frac{1}{\tJ\log(d_{\cK})}\,.
\end{align}
Here, $g_{\rm YM}$ is the classical Yang-Mills coupling \eqref{YM-coupling}.
Assuming these two conditions, we can approximate
\small
\begin{align}
    V_{\textnormal{eff}}&\approx V_0+V_{\textnormal{grav}}+V_{S^2_{\tJ}-\cK}\nn\\
    &\approx\Delta_{\cK}^4\frac{\tJ d_\cK}{g^2_{\rm YM}\,\rho^4}-\frac{\tJ^2\Delta_{\cK}^2}{R^2\rho^{4}}d_{\cK}^{3/2}+\frac{\tJ^2\,d_{\cK}}{\ell_p^2R^2\rho^{4}}\Bigg[\frac{\sqrt{d_{\cK}}\Delta_{\cK}^2}{\Delta_{\tJ}^2+\sqrt{d_{\cK}}\Delta_{\cK}^2}-\log\Big(\frac{\Delta_{\tJ}^2+\sqrt{d_{\cK}}\Delta_{\cK}^2}{\Delta_{\tJ}^2+\Delta_{\cK}^2}\Big)\Bigg]\,.
\end{align}
\normalsize
Then, the equation that describes the minimum, i.e. $\p V_{\rm eff}/\p \Delta_{\cK}\overset{!}=0$, reads
\begin{align}\label{eq:DONOToversimplify}
    \frac{\tJ^3\,d_{\cK}\Delta_{\tJ}^2  \left(\Delta_{\tJ}^2-d_{\cK} \Delta_{\cK}^{2}\right)}{R^4 \left(\Delta_{\tJ}^2+\Delta_{\cK}^{2}\right) \left(\Delta_{\tJ}^2+\sqrt{d_{\cK}} \Delta_{\cK}^{2}\right)^2}-\frac{\tJ d_{\cK}^{3/2} }{R^2}+\frac{2 \Delta_{\cK}^{2}d_\cK }{g^2_{\rm YM}}\approx 0\,.
\end{align}   
Note that the second term  arises from $V_{\rm grav}$.
Since \eqref{DeltaK-J-hierarchy} implies    $\Delta_{\cK}^2 \gg \frac{\Delta^2_{\tJ}}{d_{\cK}}$, this reduces to 
\begin{align}
     -\frac{\tJ^3 d_{\cK} \Delta_{\tJ}^2 }{R^4 \Delta_{\cK}^4}-\frac{\tJ d_{\cK}^{3/2} }{R^2}+\frac{2 \Delta_{\cK}^2 d_\cK}{g^2_{\rm YM}}\approx 0 \ .
\end{align}
We shall consider the "weak gravity`` regime where we can neglect the  contribution from $V_{\rm grav}$ (i.e. the second term). This amounts to the requirement $V_{\rm grav}\ll V_{S^2_{\tJ}-\cK}$ or equivalently
\begin{align}\label{eq:J-K-rel}
    \frac{\Delta_{\tJ}^4}{\Delta_{\cK}^4}\gg \frac{\Delta_\tJ^2R^2d_\cK^{\frac{1}{2}}}{\tJ^2}\,.
\end{align}
We then find a non-trivial solution for the minimum
\begin{align}\label{eq:DK-DJ-rel}
   \boxed{  \Delta_{\cK}^*
    \approx \left(\frac{\tJ^3 g_{\rm YM}^2 \Delta_\tJ^2}{R^4}\right)^{1/6}
    =\left(\frac{\tJ^2 \tilde{g}_p^2 \Delta_\tJ^2}{R^4\rho^2}\right)^{1/6} \ .
     }
\end{align} 
This is compatible with the simplifying  assumption \eqref{eq:1st-constraint-DK} provided 
\begin{align}\label{existence}
g_{\rm YM}^2 \Delta_\tJ^2R^2 \equiv
 \frac{ \tilde{g}_p^2 \Delta_\tJ^2R^2}{\tJ\rho^2} \gg 1\,.
\end{align}
The hierarchy requirement $\Delta_{\tJ}\ll \Delta_{\cK}$ amounts to 
\begin{align}
    \tJ^3 g_{\rm YM}^2 &\gg \Delta_{\tJ}^4 R^4\,,
\end{align}
which requires $\tJ$ to be large. 
The stability condition \eqref{eq:2nd-constraint-DK} together with \eqref{existence} requires 
\begin{align}
    \frac{1}{\Delta_\tJ^2 R^2} &\ll g^2_{\rm YM}  \ll\frac{1}{\tJ\log(d_{\cK})}\, 
\end{align}
which is consistent for
  \begin{align}  
  \Delta_\tJ^2 \gg 
   \frac{\tJ\log(d_{\cK})}{R^2} \ .
\end{align} 
The point of this analysis is not to determine the "true`` range of parameters, but to demonstrate that there are non-trivial solutions,
which may lead to a large hierarchy of scales  under suitable circumstances. 
Since the detailed form of $\Delta_\tJ$ depends sensitively on the details of the background, more work is required before making specific predictions.

As a result, the effective Newton coupling constant  \eqref{eq:G-Newton-constant} is proportional to
\begin{align}
    G_N\approx \frac{\rho^2}{\tJ^2d_{\cK}^{3/2}(\Delta_{\cK}^*)^2}\sim \frac {\rho^2}{(\tJ \sqrt{d_k})^3}\left(\frac{R^2}{g_{\rm YM}\Delta_\tJ}\right)^{\frac{2}{3}}\,.
\end{align}
As a reminder, we note again that 
the resulting Planck mass
 follows the same evolution as the effective masses \eqref{eq:effective-evolution} during the expansion of the universe. In this sense, the local physics can be considered as time-independent.

\paragraph{Kinetic term 
for $\Delta_\cK^*$.} 


Finally, 
since the minimum $\Delta_{\cK}^*$ is a function of the time-like parameter $\tau$ through $\rho$, we should also consider its kinetic contribution. The kinetic action of $\Delta_{\cK}^*$ arise from the 
mixed components $\cF_{\dot\mu\ib}$  of the $10d$ flux $\cF_{\Ibold\Jbold}$, which we have assumed to be negligible, i.e. $\cF_{\dot\mu\ib}\approx0$.
Indeed, the kinetic action for  $\Delta_{\cK}$ assuming $|\cK|=4$ takes the form 
\begin{align}
    S_{\rm mixed} \sim -\frac{1}{g^2\ell_p^4}\int\limits_{\cM^{1,3}} \Omega\, \tr_{\cH_{\cK}}\big(\{t^{\dot\mu},t^{\ib}\}\{t_{\dot\mu},t_{\ib}\}\big)
     \sim -\frac{J d_\cK^{3/2}}{g^2\ell_p^4}\int\limits_\cM \sqrt{G} \,G^{\mu\nu}\p_{\mu} \Delta_\cK\p_{\nu} \Delta_\cK
\end{align}
recalling that $t_{\ib}=m_{\cK}z_{\ib}=\Delta_{\cK}d_{\cK}^{1/4}z_{\ib}$ with $z_{\ib}z^{\ib}\sim \cO(1)$. 
To compare this with the classical action for $\Delta_\cK$
\begin{align}
S_0 \sim -\frac{\tJ}{g^2\ell_p^4}\int\limits_{\cM^{1,3}} \Omega\,\tr_{\cH_{\cK}}\big([t^{\ib},t^{\jb}][t_{\ib},t_{\ib}]\big)\,,
\end{align}
it suffices to compare the flux $\cF^{\ib\jb} \sim \{t^{\ib},t^{\jb}\} \sim \Delta_\cK^2$ on $\cK$ with the mixed flux
\begin{align}
\cF^{\dot\mu\ib} \sim \
\{t^0,t^{\ib}\} \sim \{t^0,x^4\} \frac{t^{\ib}}{x^4} \sim \frac{t^{\ib}}{R}  \frac{x^0}{x^4}\sim \frac 1R d_\cK^{1/4} \Delta_\cK
\end{align}
using \eqref{eq:DK-DJ-rel}. We observe that the mixed contribution, i.e. the kinetic term, is negligible if  
\begin{align}
    \Delta_\cK > \frac {d_\cK^{1/4}}R  \ .
\end{align}
If this "adiabatic condition``  is not satisfied, then the time-dependent vacuum would be somewhat lagging behind the minimum of the potential \eqref{eq:DK-DJ-rel}.
In fact, one may even consider rotating $\cK$ 
with the cosmic expansion along the lines of  \cite{Steinacker:2014eua} while preserving $SO(1,3)$, which may stabilize $\cK$ even classically.
Such refinements are left for future work.

\section{UV finiteness at one loop and subleading terms} \label{sec:higher-derivatives}

This section studies the UV convergence of the one-loop effective action in terms of the $\alpha$-expansion of $\sQ_{10}$. We find that all contributions are  UV finite, except for one 4-derivative term $\delta \cF_{\dot\mu\dot\nu}^4$ at order $\cO(\alpha^4)$, which formally has a logarithmic divergence. Nevertheless, this divergence is argued to be an artifact, and is regularized upon taking the fuzziness of quantum spacetime into account, cf. \cite{Steinacker:2023myp}. At the end of this section, we also discuss the hierarchy between different sectors in the one-loop effective action, whose technical details are relegated to Appendix \ref{app:hierarchy}.

\paragraph{UV finiteness at one loop.} 
Let us expand the $SO(1,9)$ character up to order $\cO(\a^6)$ 
\begin{align}\label{eq:Q10-expansion}
    \sQ_{10}=\sX_6&+\frac{\alpha^4}{4}\Big(V_{\cM}[4]+V_{\cM-\cK}[4]\Big)-\frac{\alpha^6}{24}\Big(V_{\cM}[6]+V_{\cM-\cK}[6]\Big)+\cO(\alpha^8)\,.
\end{align}
\normalsize
Here, $\sX_6$ contains only lowest-derivative contributions, while $V_{\cM}[n\geq 4]$ and $V_{\cM-\cK}[n\geq 4]$ represent higher-derivative contributions, which can be handled with the master $k$-integral \eqref{eq:general-k-integral}. Explicitly, 
the $4th$ order potentials are
\begin{subequations}
    \begin{align}
        V_{\cM}[4]&=4\delta\cF^{\dot\mu\dot\nu}\delta\cF_{\dot\nu\dot\rho}\delta \cF^{\dot\rho\dot\sigma}\delta \cF_{\dot\sigma\dot\mu}-\delta \cF^{\dot\mu\dot\nu}\delta\cF_{\dot\mu\dot\nu}\delta\cF^{\dot\rho\dot\sigma}\delta \cF_{\dot\rho\dot\sigma}\,,\label{eq:V-M-4}\\
        V_{\cM-\cK}[4]&=-2\delta \cF^{\dot\mu\dot\nu}\delta \cF_{\dot\mu\dot\nu}\delta \cF^{\ib\jb}\delta \cF_{\ib\jb}\,,\label{eq:V-M-K-[4]}
    \end{align}
\end{subequations}
and
\begin{subequations}
    \begin{align}
        V_{\cM}[6]&=-(\delta\cF^{\dot\mu\dot\nu}\delta\cF_{\dot\mu\dot\nu})V_{\cM}[4]\,,\\
        V_{\cM-\cK}[6]&=-2(\delta\cF^{\dot\mu\dot\nu}\delta\cF_{\dot\mu\dot\nu})^2(\delta\cF^{\ib\jb}\delta\cF_{\ib\jb})+(\delta\cF^{\dot\mu\dot\nu}\delta\cF_{\dot\mu\dot\nu})V_{\cK}[4]\nn\\
        &\qquad \qquad +(\delta\cF^{\ib\jb}\delta\cF_{\ib\jb})V_{\cM}[4]-2(\delta\cF^{\dot\mu\dot\nu}\delta\cF_{\dot\mu\dot\nu})(\delta\cF^{\ib\jb}\delta\cF_{\ib\jb})^2\,.
    \end{align}
\end{subequations}
\normalsize
are the $6th$ order potentials.
Here, 
\begin{align}
    V_{\cK}[4]=4\delta \cF^{\ib\jb}\delta \cF_{\jb\kb}\delta \cF^{\kb\mb}\delta \cF_{\mb\ib}-\delta \cF^{\ib\jb}\delta \cF_{\ib\jb}\delta \cF^{\mb\nb}\delta\cF_{\mb\nb}\,.
\end{align}
We will show below that $V_{\cM}[6]$ and $V_{\cM-\cK}[6]$ are  UV finite. However, there is a subtlety at the leading order $\alpha^4$ with the contribution coming from $V_{\cM}[4]$ cf. \eqref{eq:V-M-4}, which  appears to be UV log-divergent. The one-loop contribution associated with this term is
\begin{align}
    \Gamma[V_{\cM}^{(4)}]=-\frac{\im}{2}\int_0^{\infty}d\alpha\,\alpha^3\Tr_{\End(\cH_{\tJ})\otimes \mg}\Big[e^{-\im\alpha\,\Box}V_{\cM}[4]\Big]\,.
\end{align}
In the semi-classical regime, it can be shown that
\small
\begin{align}
    V_{\cM}[4]=V[\cM]^{\alpha\beta\gamma\delta}\p_{\alpha}\p_{\beta}\p_{\gamma}\p_{\delta}\mapsto V[\cM]^{\alpha\beta\gamma\delta}k_{\alpha}k_{\beta}k_{\gamma}k_{\delta}
\end{align}
\normalsize
upon evaluating on a plane-wave basis, where 
\begin{align}
    V[\cM]^{\alpha\beta\gamma\delta}\approx 4\cT^{\dot\mu\dot\nu\alpha}\cT_{\dot\nu\dot\rho}{}^{\beta}\cT^{\dot\rho\dot\sigma\gamma}\cT_{\dot\sigma\dot\mu}{}^{\delta}-\cT^{\dot\mu\dot\nu\alpha}\cT_{\dot\mu\dot\nu}{}^{\beta}\cT^{\dot\rho\dot\sigma\gamma}\cT_{\dot\rho\dot\sigma}{}^{\delta}\,,\qquad \cT^{\dot\mu\dot\nu\alpha}:=\{\{t^{\dot\mu},t^{\dot\nu}\},x^{\alpha}\}\,.
\end{align}
Using \eqref{eq:general-k-integral}, we can evaluate the $k$-integral to
\begin{align}
    \int\frac{d^4k}{(2\pi)^4\sqrt{G}}k_{\alpha}k_{\beta}k_{\gamma}k_{\delta}e^{-\im \a\, k_{\mu}k_{\nu}\rho^2G^{\mu\nu}}=+\frac{\im}{(4\pi)^2}\frac{1}{\alpha^4\rho^4}\frac{\gamma_{(\alpha\beta}\gamma_{\gamma\delta)}}{4}\,.
\end{align}
As a result,
\begin{align}
    \Gamma[V_{\cM}^{(4)}]\approx+\frac{1}{8(4\pi)^2}\int\limits_{\cM^{1,3}} \frac{\sqrt{G}}{\rho^4}\int_0^{\infty}\frac{d\a}{\a}\Tr_{\mg}\Big(e^{-\im\a\,(\Box_6+\Delta_{\tJ}^2)}V[\cM]^{\mu\nu\rho\sigma}\gamma_{(\mu\nu}\gamma_{\rho\sigma)}\Big)\,,
\end{align}
dropping the $\Tr_\hs$ for simplicity. The integral over $\a$ suffers from a logarithmic UV divergence
at $\a \to 0$, which stems from a logarithmic divergence of the momentum space integral over $k$. 
We can hence evaluate this integral formally with a UV cutoff as 
\begin{align}\label{eq:log-div-VM}
\Tr_{\mg}\Big(\log\Big[\frac{\Box_6+\Delta_{\tJ}^2}{\Lambda_{\rm UV}^2}\Big]\Big) \ .
\end{align}
As observed in \cite{Steinacker:2023myp}, the UV completion of this integral via string modes beyond the UV cutoff $\L_{\rm UV} \sim \cO(L_{\rm NC}^{-1})$ contribute to the non-local part of the effective action, as $\Box_6\mapsto m_{\cK}^2(x-y)^2+\Delta_{\cK}^2$.
Then \eqref{eq:log-div-VM} is replaced by a finite number, leading to some 4-derivative correction term to the gravitational effective action.



It is worth noting that $V_{\cM}[4]$ is the only contribution suffering from such a (formal) UV  divergence in the $\a$-expansion of $\sQ_{10}$. Indeed, at order $\alpha^{2n}$, $V_{\cM}[2n]$ comes with $k_{\alpha_1}\ldots k_{\alpha_{2n}}$ in the semi-classical regime in consideration so that the $k$-integral can be evaluated 
as\footnote{See Appendix \ref{app:A} for more explicit expression.}
\begin{align}
    \int d^4k\, (k\cdot k)^{n}e^{-\im \alpha\, k\cdot k}\sim \frac{1}{\alpha^{n+2}}\,.
\end{align}
As a consequence, we end up with the following $\alpha$-integral
\begin{align}\label{eq:VM-order-n-alpha-integral}
    \int_0^{\infty} d\alpha\, \alpha^{n-3} \Tr_{\hs\otimes \mg}\Big[e^{-\im \alpha\, (\Box_6+\Delta_{\tJ}^2)}V[\cM]^{\mu(2n)}\gamma_{\mu\mu}\ldots\gamma_{\mu\mu}\Big]\,,
\end{align}
where $\gamma_{\mu\mu}\gamma_{\mu\mu}=\gamma_{(\mu_1\mu_2}\gamma_{\mu_3\mu_4)}$, etc. Here,
\begin{align}
    V[\cM]^{\alpha(2n)}:=\cO\Big(\Tr(\cT_{\dot\mu\dot\nu}{}^{\alpha})^{2n}\Big)
\end{align}
denotes the tensorial structures resulting from taking appropriate traces over the torsions $\cT_{\dot\mu\dot\nu}{}^{\alpha}$. The  integral over $\a$ is now  convergent for $n\geq 3$.

\paragraph{Higher-order terms.} It can be shown that the $\a$-integrals associated with the mixed $\cM-\cK$ terms $V_{\cM-\cK}[n\geq 4]$ and the pure $\cK$ terms $V_{\cK}[n\geq 4]$ in  \eqref{eq:Q10-expansion}, \eqref{eq:Q10} are  convergent and UV finite, by simple power counting of $k$ and $\alpha$. Moreover, the higher-order terms in the $\alpha$-expansion of $\sQ_{10}$ are {\em subleading} compared to the lowest-order terms at $\cO(\alpha^4)$. Therefore, the 
gravitational
one-loop effective action is essentially captured by the lowest-order terms, which justifies the results of Section \ref{sec:4}. 
This is elaborated in Appendix \ref{app:hierarchy}.



\section{Discussion}\label{sec:discussion}

The main contribution of this paper is a suitable formalism to compute and interpret the one-loop effective action for the higher-spin theory, which is induced by the IKKT matrix model on the background brane $\cM^{1,3}\times \cK$. This one-loop action is UV-finite, thanks to maximal supersymmetry and the finiteness of the numbers of dof. per unit volume\footnote{This is quite different with standard higher-spin gravities, see e.g. \cite{Giombi:2013fka,Giombi:2014iua,Gunaydin:2016amv}, where higher-spin symmetry plays the central role in softening possible UV divergence. }. Computing the effective action for the induced gravity sector, we establish a clear hierarchy between lowest-order and higher-order contributions in $(\cT_{\dot\mu\dot\nu}{}^{\alpha},\delta\cF_{\ib\jb})$, as discussed in Section \ref{sec:higher-derivatives} and Appendix \ref{app:hierarchy}. The lowest-order contributions in $(\cT_{\dot\mu\dot\nu}{}^{\alpha},\delta\cF_{\ib\jb})$ turn out to be dominant, which justify previous treatment at one loop, cf. \cite{Steinacker:2021yxt,Steinacker:2023myp,Steinacker:2024huv}. This allows  to compute the effective Newton constant or Planck mass $\frac{1}{G_N}\sim \Delta_{\cK}^2\rho^{-2}$, which turns out to follow the same evolution as the effective KK mass \eqref{eq:effective-evolution} along the cosmic expansion. This is an important result, since it allows the local physics to be independent of the cosmic evolution.

Another important topic is the stabilization of the extra dimensions $\cK$, or at least of its radius. Refining previous considerations, we are able to give some explicit formulas for the 
KK mass scale $m^2_\cK$. This arises from a non-trivial minimum of the 1-loop effective action, which provides a UV mass scale and breaks conformal symmetry.
In this calculation, the role of the $\hs$ mass scale $\Delta_{\tJ}$ is crucial, as noticed earlier in \cite{Steinacker:2024huv}. 
The non-trivial vacuum under consideration induces a non-vanishing $\hs$ mass scale  $\Delta_{\tJ}\neq 0$ in the loops.
However, $\Delta_\tJ^2$ depends sensitively on the background  brane $\cM^{1,3}$. In particular, we found a negative value for $m_s^2$, cf. \cite{Steinacker:2023cuf}. However, as noted above, that calculation should be taken with a grain of salt. In fact, we show in Appendix C how this may be avoided on modified  backgrounds. 

An important question which remains to be clarified in future work is whether all $\hs$ masses are in the IR regime with a large hierarchy $\Delta_\cK \gg \Delta_\tJ$,
or only the graviton remains in the IR regime while other higher-spin modes acquire a large mass. In any case,
the present mechanism for stabilizing $\cK$ will work for a large range of $\Delta_\tJ$, within some bounds. 

\begin{figure}[ht!]
    \centering
    \includegraphics[scale=0.78]{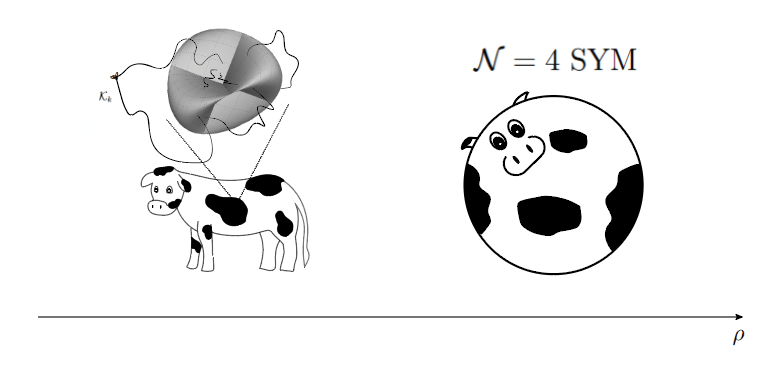}
    \caption{For some regime of $\rho$,
     non-trivial vacua can arise with stabilized $\cK$, which breaks conformal symmetry and gives rise to interesting physics. This is depicted with a $\hs\otimes \mg$-valued cow, arising from  some compact brane $\cK$ in the extra dimensions. On the other hand, at late times $\cK$ may shrink to a point, and the $\hs\otimes\mg$-valued cow becomes a spherical cow as conformal symmetry is effectively restored.}
    \label{fig:evolution}
\end{figure}

Furthermore,  our mechanism to stabilize $\cK$ is only valid in some regime where $\rho$ is not too large. At very late times, i.e. when $\rho$ gets too large, it may happen that $\Delta_{\cK}\rightarrow 0$ and $\cK$ will shrink to a point, at least in the present approximation. In that case, it is possible that $\hs$-IKKT will become a higher-spin extension of $\cN=4$ SYM without an explicit KK scale, see Fig \ref{fig:evolution}. 
However, taking into account time-dependent deformations of the space-time background $\cM^{1,3}$ along the lines of appendix \ref{app:verify} will modify $\rho$ and may prevent it from growing too large; this will be investigated elsewhere.
Other mechanisms for stabilizing $\cK$ are also conceivable, and this issue requires more work.

 Of course, interesting physics can only occur when $\Delta_{\cK}>0$ so that $\cK$ can induce masses to all (higher-spin) fields, as discussed in \cite{Steinacker:2024huv}.
This would lead to an explicit
 theory for massive higher-spin fields 
where all vertices can be derived by averaging over $S^2_{\tJ}$. This would be an interesting and vast field for further studies. In particular,
it would be interesting to compare  the interacting vertices of $\hs$-IKKT with recent attempts of constructing Lorentz-invariant vertices of massive fields in e.g. \cite{Metsaev:2005ar,Metsaev:2022yvb,Buchbinder:2022lsi,Skvortsov:2023jbn}. 


\section*{Acknowledgement}
Interesting discussions with Alexander Frenkel are gratefully acknowledged. 
HS would like to thank Pei-Ming Ho and Hikaru Kawai for useful discussions 
and hospitality at NTU Taiwan,  Jun Nishimura and Asato Tsuchiya  for discussions and hospitality at KEK and Shizuoka University, as well as Chong-Sun Chu for disussions.
This work is supported by the Austrian Science Fund (FWF) grant P36479.

\appendix
\section{Supplementary material}\label{app:0}
\subsection{Useful relations}
This appendix provides additional relations for $\cM^{1,3}$ used in the main text. All the relations here are extracted from \cite{Sperling:2018xrm,Steinacker:2019awe,Sperling:2019xar}. Let $x_a$ for $a=0,1,2,3,4$ be coordinates of $H^4$ where $x_ax^a=-R^2$ and $x^a$ transform as vectors under $SO(1,4)$ with generator $m^{ab}$. After projecting out $x^4=R\sinh(\tau)$, we have the following relations
\begin{subequations}
\begin{align}
    x_{\mu}\ttb^{\mu}&=0\,,\qquad \qquad \qquad \qquad \quad \quad \qquad \qquad \qquad\quad \ \  \mu=0,1,2,3\,, \label{eq:orthogonalofPY}\\
    \eta_{\mu\nu}\ttb^{\mu}\ttb^{\nu}&=\frac{1}{\ell_p^2}+\frac{y_4^2}{\ell_p^2R^2}=+\ell_p^{-2}\cosh^2(\tau)\,, \qquad \qquad \quad \ \eta_{\mu\nu}=\diag(-,+,+,+)\,, \label{S2sphereM13}\\
   x^\mu x_\mu &= - R^2 \cosh^2(\tau) 
   \label{yy-square} \\ 
     \{\ttb^{\mu},x^{\nu}\}&=+\frac{\eta^{\mu\nu}}{R}x^4=\eta^{\mu\nu}\sinh(\tau)\,,\\
     \{x^4,x^{\mu}\}&=\ell_p^2R\ttb^{\mu}\,,\\
    \{\ttb^{\mu},x^4\}&=-\frac{x^{\mu}}{R}\,,\\
    \{\ttb^{\mu},\ttb^{\nu}\}&=-\frac{1}{\ell_p^2R^2}\theta^{\mu\nu}\,,\\
    m^{\mu\nu}&=-\frac{\theta^{\mu\nu}}{\ell_p^2}=-\frac{1}{\cosh^2(\tau)}\Big(\sinh(\tau)(x^{\mu}\ttb^{\nu}-x^{\nu}\ttb^{\mu})+\eps^{\mu\nu\sigma\rho}x_{\sigma}\ttb_{\rho}\Big)\label{mgenerator}\,,\\
    \ttb_{\mu}\theta^{\mu\nu}&=-\sinh(\tau)x^{\nu}\,,\\
    x_{\mu}\theta^{\mu\nu}&=-R^2\sinh(\tau)\ttb^{\mu}\,,\\
    \theta_{\mu}{}^{\alpha}\theta^{\mu\beta}&=R^2\ell_p^2\eta^{\alpha\beta}-R^2\ell_p^4\,\ttb^{\alpha}\ttb^{\beta}+\ell_p^2\,x^{\alpha}x^{\beta}=R^2\ell_p^2\eta^{\alpha\beta}-R^2\ell_p^2\cosh^2(\tau)u^{\alpha}u^{\beta}+\ell_p^2\,x^{\alpha}x^{\beta}\,.
\end{align}
\end{subequations}
that are frequently used in the main text where $\ttb$ is the momentum background. Note that in the local physical regime, the Poisson structures $\theta^{\mu\nu}$ are simplified to
\begin{align}\label{flat-theta}
    \theta^{\mu\nu}\approx \frac{\ell_p^2\sinh(\tau)}{\cosh^2(\tau)}(x^{\mu}\ttb^{\nu}-x^{\nu}\ttb^{\mu})= \frac{\ell_p\sinh(\tau)}{\cosh(\tau)}(x^{\mu}u^{\nu}-x^{\nu}u^{\mu})\,,\qquad u^{\mu}=\frac{\ell_p}{\cosh(\tau)}\ttb^{\mu}\,.
\end{align}
Namely, we drop all terms associated to the Levi-Civita symbols $\epsilon^{\mu\nu\rho\sigma\delta}$ since they are subleading. Finally, one can also easily confirm the following relations
\begin{align}
    \{x^4,x^{\mu}\}=\ell_p^2R\ttb^{\mu}\,,\qquad \{x^4,\ttb^{\mu}\}=\frac{x^{\mu}}{R}\,.
\end{align}
\subsection{Useful integrals}\label{app:A}

In the main text, we often work with the following oscillatory Gaussian $k$-integrals 
\begin{align}
    \int \frac{d^4k}{(2\pi)^4\sqrt{G}}k_{\alpha_1}\ldots k_{\alpha_{2n}}e^{-\im \alpha\, k\cdot k}= \int \frac{d^4k\,(k\cdot k)^{n}}{(2\pi)^4\sqrt{\gamma}}\frac{\cG_{\alpha(2n)}}{\rho^4}e^{-\im \alpha k_{\mu}k_{\nu}\gamma^{\mu\nu}}\,,
\end{align}
where $k\cdot k = k_\mu k_\nu \gamma^{\mu\nu}$ and  $\gamma_{\mu\nu}=\rho^{-2}G_{\mu\nu}$ is considered as constant, cf. \eqref{eq:eff-G}. In particular, 
$\sqrt{\gamma}=\rho^{-4}\sqrt{G}$. Note also that
\begin{align}
    \cG_{\alpha(2n)}=\frac{\gamma_{\alpha\alpha}\ldots\gamma_{\alpha\alpha}}{4(4+2)\ldots(4+2n)}=\frac{1}{\rho^{2n}}\frac{G_{\alpha\alpha}\ldots G_{\alpha\alpha}}{4(4+2)\ldots (4+2n)}\,,\qquad n\in \N^+\,.
\end{align}
Here, the same indices mean symmetrization, e.g. 
\begin{align}
    G_{\alpha\alpha}G_{\alpha\alpha}=G_{\alpha_1\alpha_2}G_{\alpha_3\alpha_4}+G_{\alpha_1\alpha_3}G_{\alpha_2\alpha_4}+G_{\alpha_1\alpha_4}G_{\alpha_2\alpha_3}\,.
\end{align}
Of course, these oscillatory $k$-integrals are 
ill-defined as they stand.
As discussed following \eqref{duhamel-identity},
they are defined\footnote{
This may also be justified by first integrating over the Schwinger parameter and then computing the trace, which is regularized by the $\im\varepsilon$. However, the present Schwinger formalism provides more elegant closed formulas.}
by giving the Schwinger parameter $\a \to \a - i \varepsilon$ a small negative imaginary part.
Then,
\begin{align}
    \int \frac{d^4k}{(2\pi)^4}(k\cdot k)^n e^{-\im\alpha\,k\cdot k}\,
\end{align}
can be evaluated 
by applying a Wick rotation $\int dk_0\mapsto \im \int dk_4$. At this stage, we can rescale $k_{\mu}\mapsto \sinh^{-1}(\tau)k_{\mu}$. Then, effectively, we can do the integral over $k$ as if we are on flat space. Altogether,
\begin{align}\label{eq:general-k-integral}
    \int \frac{d^4k}{(2\pi)^4\sqrt{G}}k_{\alpha_1}\ldots k_{\alpha_{2n}}e^{-\im\alpha k\cdot k}&=-\frac{1}{(4\pi)^2}\frac{\im^{1-n}}{\alpha^{2+n}}\frac{1}{\rho^{4+2n}}\frac{G_{\alpha\alpha}\ldots G_{\alpha\alpha}}{2^n}\nn\\
    &=-\frac{1}{(4\pi)^2}\frac{\im^{1-n}}{\alpha^{2+n}}\frac{1}{\rho^4}\frac{\gamma_{\alpha\alpha}\ldots \gamma_{\alpha\alpha}}{2^n}
\end{align}
From the above master formula, we get for instance
\begin{align}
    \int \frac{d^4k}{(2\pi)^4\sqrt{G}} \,e^{-\im \alpha \,k\cdot k} &=  -\frac{\im}{(4\pi)^2\alpha^{2}} \frac{1}{\rho^4}\,,\nn\\
    \int \frac{d^4k}{(2\pi)^4\sqrt{G}} \, k_{\mu}k_{\nu}\, e^{-\im \alpha \,k\cdot k} &= -\frac{1}{2(4\pi)^2\alpha^{3}}\frac{\gamma_{\mu\nu}}{\rho^4}\,.
\end{align}
We can also evaluate the $\alpha$-integral as 
\begin{equation}
    \int\e^{-\im\alpha\,( \Box_6+\Delta_{\tJ}^2)}\alpha^{n}\diff\alpha =\frac{(-1)^{n+1}\im^{n+1}n!}{(\Box_6+\Delta_{\tJ}^2)^{n+1}}\,,\qquad n\in \N^+\,.
\end{equation}


\section{Hierarchy}\label{app:hierarchy}
This appendix shows that contributions associated with higher-order $\a$ expansion from $\sQ_{10}$ are generally sub-leading. This justifies the treatment of  the gravitational one loop effective action
in previous work \cite{Steinacker:2021yxt,Steinacker:2023myp,Steinacker:2024huv} 
at leading order. 
There are three classes of contributions we want to investigate:
\begin{align}
    \Gamma[V_{\cM}^{(n)}]\,,\qquad \Gamma[V_{\cM-\cK}^{(n)}]\,,\qquad \Gamma[V_{\cK}^{(n)}]\,,
\end{align}
where $\Gamma[V_{\bullet}^{(n)}]$ is the contributions to the one-loop effective action of the potential $V_{\bullet}[n]$ with $n$ being the power of $\a$ in the perturbative $\a$-expansion of $\sQ_{10}$.

\paragraph{$\Gamma[V_{\cM}^{(n)}]$ class.} These contributions can be estimated by evaluating the $\a$-integral in \eqref{eq:VM-order-n-alpha-integral} explicitly. Neglecting unimportant prefactors, we obtain
\begin{align}
    \Gamma[V_{\cM}^{(2n)}]\sim \Tr_{\hs\otimes\mg}\Big(\frac{V[\cM]^{\sigma(2n)}\gamma_{\sigma\sigma}\ldots\gamma_{\sigma\sigma}}{(\Box_6+\Delta_{\tJ}^2)^{n-2}}\Big)\,,\qquad n\geq 3\,,
\end{align}
where 
\begin{align}
    V[\cM]^{\sigma(2n)}:\approx \cO\Big((\cT_{\dot\mu\dot\nu}{}^{\sigma})^{2n}\Big)\,.
\end{align}
Assuming $\Delta_{\cK}\gg \Delta_{\tJ}$, we can use the geometric trace \eqref{eq:geo-trace} to obtain
\begin{align}
    \Gamma[V^{(2n)}_{\cM}]\sim \frac{\tJ^2}{2n-4}\frac{d_{\cK}}{\Delta_{\cK}^{2n-4}}\cO\Big((\cT_{\dot\mu\dot\nu}{}^{\sigma})^{2n}\Big)\,,\qquad n\geq3
\end{align}
for large $d_{\cK}$.
These are higher-derivative contributions to gravity, which are suppressed by the UV scale $\Delta_\cK$. 
Therefore these terms do not play a significant role for long wavelengths.

\paragraph{$\Gamma[V_{\cM-\cK}^{(n)}]$ class.} Now, consider the mixed contributions $\Gamma[V_{\cM-\cK}^{(2n)}]$. At order $\cO(\a^{2n})$, we have
\begin{align}
    \Gamma[V_{\cM-\cK}^{(2n)}]=\sum_{i+j=n}\Gamma[\delta\cF_{\dot\mu\dot\nu}^{(2i)},\delta \cF_{\ib\jb}^{(2j)}]
\end{align}
where $i,j$ denote the numbers of the operators $\delta \cF_{\dot\mu\dot\nu}$ and $\delta\cF_{\ib\jb}$ in the corresponding sub potentials $V(\delta\cF_{\dot\mu\dot\nu},\delta \cF_{\ib\jb})[i,j]$, respectively.\footnote{For $j=0$, we return to the $\Gamma[V_{\cM}^{(2n)}]$ case discussed above.} For specific values of $(i,j)$ with $i+j=n$, after taking appropriate traces and convert $\delta\cF_{\dot\mu\dot\nu}\mapsto\cT_{\dot\mu\dot\nu}{}^{\alpha}k_{\alpha}$ upon evaluating on plane wave basis,  the integral over $k$'s can be evaluated, leading to
\begin{align}\label{eq:heuristic[i,j]}
    \Gamma[(\cT_{\dot\mu\dot\nu}{}^{\alpha})^{2i},\delta\cF_{\ib\jb}^{(2j)}]&\sim \int_0^{\infty}d\a \,\a^{2j+i-3}\Tr_{\hs\otimes\mg}\Big[\cO\Big((\cT_{\dot\mu\dot\nu}{}^{\sigma})^{2i},\delta \cF_{\ib\jb}^{2j}\Big)e^{-\im\a\,(\Box_6+\Delta_{\tJ}^2)}\Big]\nn\\
    &\sim \Tr_{\hs\otimes\mg}\Big[\frac{\cO\Big((\cT_{\dot\mu\dot\nu}{}^{\sigma})^{2i},\delta \cF_{\ib\jb}^{2j}\Big)}{(\Box_6+\Delta_{\tJ}^2)^{2j+i-2}}\Big]\,.
\end{align}
We can simplify the situation by assuming that $\Delta_{\cK}\gg \Delta_{\tJ}$. Then, the trace over $\mg$ results in
\begin{align}
    \Gamma[(\cT_{\dot\mu\dot\nu}{}^{\alpha})^{2i},\delta\cF_{\ib\jb}^{(2j)}]\sim \frac{1}{d_{\cK}^{\frac{2j+i-2}{2}}}\frac{1}{\Delta_{\cK}^{2i-4}}\Tr_{\hs}\Bigg[\cO\Big((\cT_{\dot\mu\dot\nu}{}^{\sigma})^{2i}\Big)\Bigg]\,.
\end{align}
In particular, all higher-order corrections associated to the case $i=1$ and $j\geq 2$, which contribute to the induced Einstein-Hilbert action \eqref{grav-action-leading-expand}  are clearly suppressed by a factor $d_{\cK}^{\frac{2-i-2j}{2}}$. 
This means that the results in section \ref{sec:4} 
can be trusted.

\paragraph{$\Gamma[V_{\cK}^{(n)}]$ class.} This is a special case of $\Gamma[V_{\cM-\cK}^{(2n)}]=\sum_{i+j=n}\Gamma[\delta\cF_{\dot\mu\dot\nu}^{(2i)},\delta \cF_{\ib\jb}^{(2j)}]$ where we have $i=0$ and $j=n$. From \eqref{eq:heuristic[i,j]}, we get
\begin{align}
    \Gamma[V_{\cK}^{(2n)}]&\sim \int_0^{\infty}d\a \,\a^{2n-3}\Tr_{\hs\otimes\mg}\Big[\cO\Big(\delta \cF_{\ib\jb}^{2n}\Big)e^{-\im\a\,(\Box_6+\Delta_{\tJ}^2)}\Big]\
    \sim \tJ^2\,\frac{\Delta_{\cK}^4}{d_{\cK}^{n-1}}\,.
\end{align}
These are higher-order contributions to the potential \eqref{eq:potential-for-K}, which are again suppressed by 
$\frac 1{d_\cK^{n-1}}$.


\paragraph{Remark.} 
The above considerations can be summarized as follows. All higher-order contributions in $\delta\cF_{\ib\jb}$ 
to some given space-time effective action
(with fixed power of $\delta \cF_{\dot\mu\dot\nu}$) are sub-leading and may hence be neglected. Note that it will be also interesting to apply the same analysis to determine $\Gamma[V^{(n)}_{S^2_{\tJ}-\cK}]$ and $\Gamma[V^{(n)}_{S^2_{\tJ}}]$. We leave this study for a future work.

\section{On the $\hs$ scale $\Delta_\tJ$ and positive $\hs$ masses}\label{app:verify}
This appendix shows that $m_s^2$ can be positive for some suitable background. To set the stage, we consider the following modified background (cf. \cite{Battista:2023glw})
\begin{align}
    T^{\dot\mu}=f(\tau)\ttb^{\dot\mu}
\end{align}
in the semi-classical regime,
where $f(\tau)$ is some function which preserves $SO(1,3)$. Then,
\begin{align}
    \{T^{\dot\mu},\{T_{\dot\mu},u^{\nu(s)}\}\}=s(s-1)\{T^{\dot\mu},u^{\nu}\}\{T_{\dot\mu},u^{\nu}\}u^{\nu(s-2)}+s\{T^{\dot\mu},\{T_{\dot\mu},u^{\nu}\}\}u^{\nu(s-1)}\,.
\end{align}
Since $f(\tau)$ needs to preserve $SO(1,3)$, we can set
\begin{align}
    f(\tau)=f\Big(-\frac{x^2}{R^2}\Big)=f(\cosh(\tau)) \ .
\end{align}
With hindsight, we consider
\begin{align}
    f(\cosh(\tau))=\frac{1}{\cosh(\tau)}\,
\end{align}
 and use $\cosh(\tau) \sim \sinh(\tau)$ at late times.
Then, it can be shown that
\small
\begin{align}
    \{T^{\dot\mu},T^{\nu}\}
    =-\frac{1}{R^2\cosh^3(\tau)}\epsilon^{\dot\mu\nu\sigma\rho}x_{\sigma}u_{\rho}\,.
\end{align}
\normalsize
Furthermore, one finds
\begin{align}
    \{T_{\mu},\{T^{\mu},u^{\nu}\}\}
    &=-\frac{2}{R^2\cosh^4(\tau)}u^{\nu}\,.
    \label{TTu-brackets}
\end{align}
Putting this together gives
\begin{align}
    \{T_{\dot\mu},\{T^{\dot\mu},u^{\nu(s)}\}\}\approx-\frac{s(s+1)}{R^2\cosh^4(\tau)}u^{\nu(s)}\,
\end{align}
 at late times.
Thus, on the background $T^{\dot\mu}=\frac{1}{\cosh(\tau)}\ttb^{\dot\mu}$, all  $\hs$ masses are positive,
\begin{align}
    m_s^2=\frac{s(s+1)}{R^2\cosh^4(\tau)} \geq 0\,.
\end{align}
As a result, the $\hs$ scale $\Delta_{\tJ}$ is positive, and can be set as
\begin{align}
    \Delta_{\tJ}^2:=\frac{\tJ}{R^2\cosh^4(\tau)}\, > 0 \ .
\end{align}
 However, such a time-dependent background entails a host of other issues (such as local normal coordinates etc.) which will be 
addressed elsewhere. Then, the analysis in Section \ref{sec:stabilize} for stabilizing $\cK$ needs some revision, but it is 
now evident that negative $\hs$ masses 
(such as $m_s^2 = - \frac{s}{R^2}$ for the $T^\mu \sim t^\mu$ background \cite{Steinacker:2023cuf})
can and should be avoided by carefully choosing backgrounds that are fully consistent.




\setstretch{0.8}
\footnotesize
\bibliography{twistor}
\bibliographystyle{JHEP-2}

\end{document}